# A conjecture on demographic mortality at high ages

**Giuseppe Alberti**
Independent Researcher
-ORCID: 0000-0002-3041-8016-
**email: giuseppe.alberti@squ-systems.eu**

**Abstract**

*The study considers the model of an abstract organism, called Arbitrary Oscillator (ArbO), which is capable of making decisions at each timed step. These decisions are 'critical' since, randomly, their outcome can be 'fatal' for ArbO, thus bringing its life cycle to an end. If we impose limits on the total number of critical decisions using a fixed parameter TC (Total Cases), we can treat the statistical distribution of fatal events over a large number of ArbOs using statistical mechanics methods. This results in a mathematically definable asymmetric 'bell' distribution, which can be compared with demographic mortality curves (dx curves), with an appropriate choice of time scale (one step = five years). The possibility of modeling and therefore predicting the trend of demographic mortality is of great scientific and social interest. Our conjecture assumes that, as demographic longevity improves, i.e., with the lengthening of lifespan, the actual demographic curves will increasingly match the mathematical distribution curve of our ArbO. The statistical distribution of the ArbO was introduced by the author in a previous paper and is here recalled and formalized analytically and its characteristics are detailed. The above said conjecture is based on two case studies: mortality in the United States from 1900 to 2017 and mortality in Italy from 1974 to 2019. The conjecture, applied to both case studies, appears reasonable. Tables and comparison figures are provided to support this. Also, an attempt to predict demographic mortality behavior and limitations for the years to come is provided. Finally, the more general theme of the nature of human aging can also be related to our conjecture, since it can highlight the presence of an absolute limit on the number of 'critical' events (the TC parameter). As 'critical' events accumulate over time by aging, approaching the final limit value, the probability of death will tend toward one.*

## 1. Introduction and recalls

Operating on demographic Life Tables, mortality curves can be obtained that illustrate the number of deaths in a given age interval when that age interval spans a lifetime. These statistical data are normally presented from year to year in tabular form for a standard number of total cases (100,000 people) and represent the evolution of life possibility to the social conditions of the demographic community under consideration. A typical generic mortality curve takes the form presented in Fig. 1. Three main highlighted areas are noted: an area A representing infant mortality, an area B representing a quasi-constant number of deaths (i.e., not very dependent on age), and finally an area C related to the more evident peak mortality at advanced ages. The sum A+B+C leads to 100000 dx events.

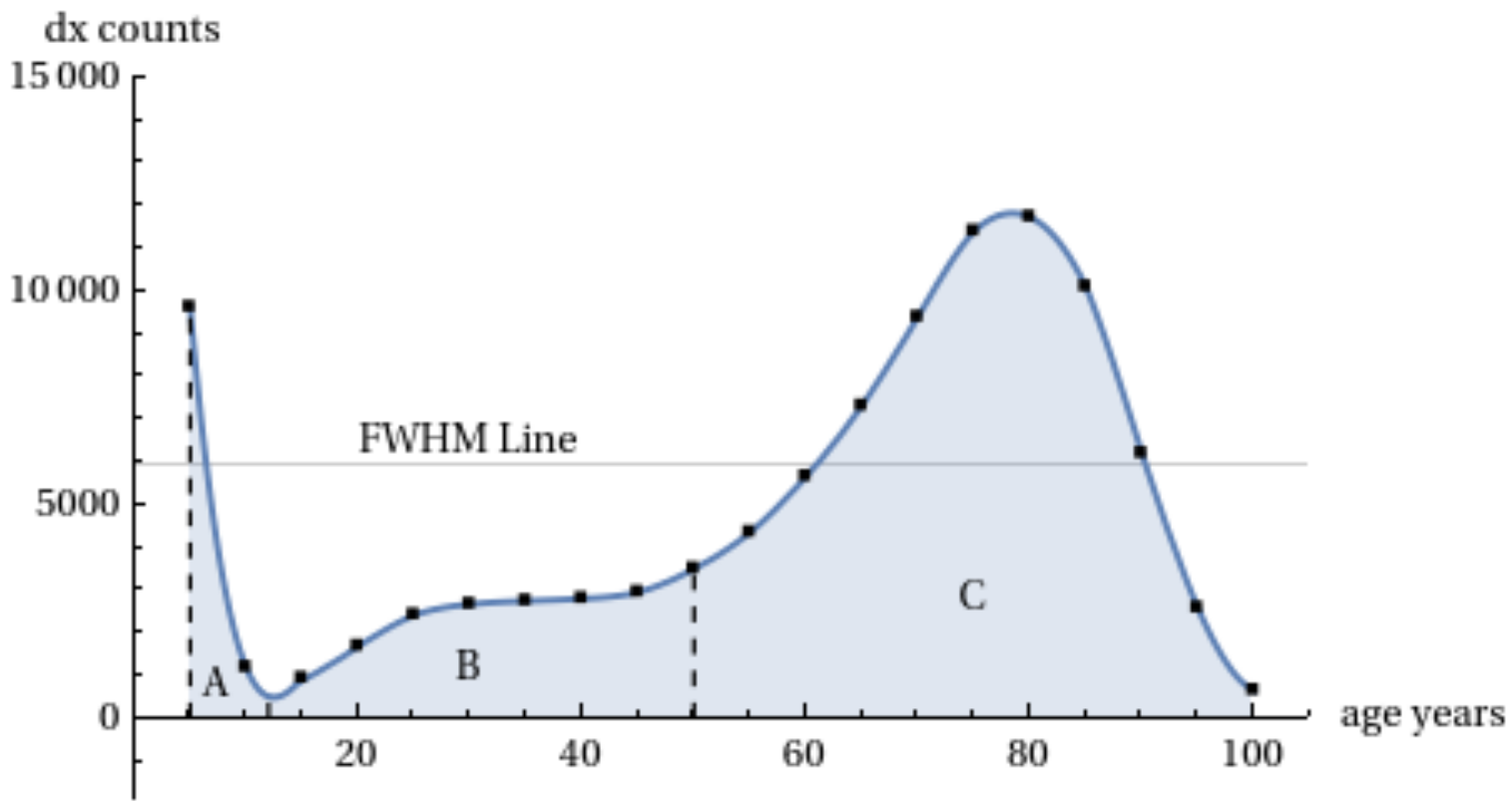

Fig.1 - Mortality poll points & interpolated curve with A,B,C areas and FWHM line along age intervals

The plotted single points corresponds to demographic mortality data i.e. the deceases (called dx counts in the Life Table) for the considered age interval (in this case five years wide). The continuous curve is the interpolation of the point data. A horizontal line is also presented useful to define the FWHM (Full Width at Half Maximum) parameter associated to the curve peak. The FWHM figure measures the age width of the segment intercepted between the FWHM line and the continuous curve. In this study we present the thesis that, as lifespan increases, the mortality curve also changes, converging to a curve of a statistical type that we refer to as the S-system curve. In other words, areas A and B will tend to cancel out, while area C will follow the pattern of an S-type curve (which itself has no significant A or B component). This curve is describable analytically as a function of a time variable (age of the person) and a parameter equal to the total area subtended by the curve. This curve formula was presented in a 2022 study as per Ref. [1] (also detailed in Ref. [2]). In that Ref. [1] work we introduced the S-system object and its properties. Correlated to this we also introduced the so called Arbitrary Oscillator (ArbO). The curve under discussion is derived by the analysis of these two objects. Hereunder we provide some recall of these concepts for the ease of the reader. In the second section we describe in more detail the characteristic of the curve while in the other following sections we deal with the mortality real demographic data and their possible convergence toward the theoretical mortality curve including a possible future mortality forecast. Finally, the conclusions test the credibility of the conjecture and outline possible areas of research development a for further confirmation of the conjecture

***S-systems mathematical definition***

The S system of equations is defined as follows:

$$\sum_{i=1}^{\text{imax}} (0.5)^i\, m_i = 1 \qquad\qquad m_i \text{ positive or null integers}$$

$$\sum_{i=1}^{\text{imax}} m_i = \text{TC}$$

$$\text{TC} = \text{imax} + 1$$

As a consequence of above equations, the solutions $\{m_i\}$ of the system must be :

$$0 \le m_i \le 2^i \qquad\qquad \text{with} \qquad 1 \le i \le \text{imax}$$

Note that the first eq. can be expressed also in the equivalent form :

$$\sum_{i=1}^{\text{imax}} 2^{\text{imax}-i}\, m_i = 2^{\text{imax}}$$

obtained by multiplying both members by $2^{\text{imax}}$. $A$ simple example with TC = 5 can clarify the system. The $S$ system in this case will be :

$$8\, m_1 + 4\, m_2 + 2\, m_3 + m_4 = 16$$
$$m_1 + m_2 + m_3 + m_4 = 5$$

and the three possible solutions are:

$$\begin{pmatrix} m_1 \to 0 & m_2 \to 3 & m_3 \to 2 & m_4 \to 0 \\ m_1 \to 1 & m_2 \to 0 & m_3 \to 4 & m_4 \to 0 \\ m_1 \to 1 & m_2 \to 1 & m_3 \to 1 & m_4 \to 2 \end{pmatrix}$$

***The Arbitrary Oscillator object***

This ArbO can be associated to a $S^{TC}$ system. We imagine indeed, just like the classic harmonic oscillator well known in physics, an object that can move between two fixed positions, but on the contrary with the classic oscillator, the "decision" to move to and from is not sure but aleatory. So the Arbitrary Oscillator, at any step of time *r* (or clock step), can stay still or jump to the other position. If we tag the "decisions" with a label "0" or "1" meaning e.g. stay or move, we can have a sequence of decisions or "cells" list like e.g {0,1,0,0,1...} that define the story of the motion and relevant "decisions". We see also that at any previous decision two next future decisions, 0 or 1, can arise. So, all around the clock time, the number of possible sequences will expand in an exponential way : $2^r$. But if we admit the possibility to have some "death" instance (again as random instance) at one or more of the possible choices available at the generic step *r,* we will have a choice count at step *r* not exponentially determined but only defined by the preceding evolution as follows:

(Equation group S) (EGS)

$$m_r + v_r = 2\, v_{r-1};\ 0 \le m_r,\ v_r \le 2^r;\ m_r,\ v_r \text{ integers};\ v_0 = 1;\ r \ge 1$$

where $m_r,\ v_r$ represent the number of mortal events and safe events respectively at the end of the *r*th step. This process can continue indefinitely or stop when at some *"rend"* step we will reach $v_{\text{rend}}=0$. At this moment the ArbO device will not have any more *v* "fuel" to generate the next step. In this case there will be a total final count of previous $m_r,\ v_r$ integer numbers and if we fix a max $m_r$ sum defined as TC ("Total Counts") we will result in a $S^{TC}$ system as shown hereunder.

***The math of ArbO***

The mathematical description of the ArbO object is conditioned to a criterion of max limit of evolution : the ArbO can't indefinitely evolve, so a limit is imposed with the condition:

$$\sum_{r=1}^{\text{Rmax}} m_r = \text{TC}$$

Coming back to the above (EGS) eq.s, we observe that these lead also to a recursive solution:

$$v_r = 2^r - \sum_{t=1}^{r} 2^{r-t}\, m_t$$

meaning that the $v_r$, as expected, depends from the evolution of the $m_r$. Imposing the existence of an Rmax such that $v_{\text{Rmax}}=0$, we will result into a system of diophantine equations like:

$$\sum_{r=1}^{\text{Rmax}} m_r = \text{TC};\ \sum_{r=1}^{\text{Rmax}} 2^{-r}\, m_r = 1;\ \text{Rmax} = \text{TC} - 1;\ \sum_{r=1}^{\text{Rmax}} v_r = \text{TC}-2;\ Q_r = m_r + v_r;\ \sum_{r=1}^{\text{Rmax}} Q_r = 2\,(\text{TC}-1);\ \text{(Equation group EGDS) (EGDS)}$$

We see that the first three eq.s of (EGDS) are the usual $S^{TC}$ system equations previously introduced. We conclude that a diophantine $S^{TC}$ system describes the ArbO object to which we can add also the $v_r$ variables (specific of the ArbO mechanism ) that, at any rate, are fully determined by the independent $m_r$ variables. For the e.g. already known example of TC=5 we have then all the possible solutions, including the $v_r$ variables:

$$8\, m_1 + 4\, m_2 + 2\, m_3 + m_4 = 16$$

$$m_1 + m_2 + m_3 + m_4 = 5$$

$$\{\{m_1 \to 0,\ m_2 \to 3,\ m_3 \to 2,\ m_4 \to 0,\ v_0 \to 1,\ v_1 \to 2,\ v_2 \to 1,\ v_3 \to 0,\ v_4 \to 0\},$$
$$\{m_1 \to 1,\ m_2 \to 0,\ m_3 \to 4,\ m_4 \to 0,\ v_0 \to 1,\ v_1 \to 1,\ v_2 \to 2,\ v_3 \to 0,\ v_4 \to 0\},$$
$$\{m_1 \to 1,\ m_2 \to 1,\ m_3 \to 1,\ m_4 \to 2,\ v_0 \to 1,\ v_1 \to 1,\ v_2 \to 1,\ v_3 \to 1,\ v_4 \to 0\}\}$$

***The ArbO configurations***

The (EGDS) eq.s not fully describe the ArbO possible evolutions sequences. The (EGDS) describe the total quantities of $m_r,\ v_r$ along the *r* steps evolution, but these do not specify on which "cell" (or decision place), from the $Q_r$ available, the effective deadly events occur. This aleatory condition holds for all the solutions of a S system, in the sense that given a particular solution, one can wonder about the number of the associated configurations ( i.e. 0,1 sequences) to this specific solution. Now considering the exponential growth of the number of solutions in the above systems, this growth is again more enhanced for the possible associated configurations. For example (Ref. [2]) with TC=21, the S system will covers 30410 different $\{m_r,\ v_r\}$ solutions. Between these solutions there will be one (or some) with the maximum no. of configurations, i.e, 14.112.000 configurations for each individual solution.

***The most probable solution***

Applying the methods of the Fermi statistics (by virtue of some formal analogies between the arbitrary oscillator model above described and the quantum models studied in the last century by E. Fermi, (Ref. [3]), one seek for the most probable solution of a ArbO - S system i.e. the solution that shows the higher no. of configurations compatible with some boundary limiting conditions. In Ref. [1] we found a recursive equation to compute this maximum likelihood solution. We also identified an analytic representation of this discrete solution with a continuous real $r$ variable. This theoretical ArbO most probable mortality $m(r)$ has the following form, with "$r$" intended as a time continuous variable and TC a parameter equal to the area of the curve:

$$m\,(r,\ \mathrm{TC}) = 2^{2\,r-\frac{(1+2\,g)\,\mathrm{Log}[2^{\mathrm{rF}}+2^{r}\,g]}{g\,\mathrm{Log}[2]}}\ \mathrm{c1} \tag{1}$$

The g, k terms are numbers while c1 and rF are constant depending only from the TC parameter, as follows, where Log is the natural logarithm function :

$$\begin{gathered} g\ =\ (1-k)\ ;\ k = \mathrm{Log}[2] \\ \mathrm{c1} = (1+g)\left(2^{\mathrm{rF}}+g\right)^{1+\frac{1}{g}}\,k \\ \mathrm{rF} = \frac{\mathrm{Log}(-2+\mathrm{TC})}{k} \end{gathered} \tag{2}$$

In according with the S-System EGDS equations, also for the continuous form $m(r,\mathrm{TC})$ the following conditions will hold:

$$\int_0^{\infty} m(r,\ \mathrm{TC})\,d\,r\ = \mathrm{TC};\ \int_0^{\infty} m(r,\ \mathrm{TC})\,2^{-t}\,d\,t\ =\ 1; \tag{3}$$

# 2. Properties of the *m*(r,TC)

If we look to all the terms in eq.s (2) we see that they can be reduced to numbers or to a functions of TC. Then, by substituting them into Eq. (1), we obtain the following expression of $m(r,\mathrm{TC})$, that we will rename as $\mathrm{mTC}(r,\mathrm{TC})$ :

$$\mathrm{mTC}\,(r,\ \mathrm{TC})\ =\ -4^{r}\,(-1+\mathrm{TC}-\mathrm{Log}[2])^{1+\frac{1}{1-\mathrm{Log}[2]}}\,(-2+\mathrm{Log}[2])\,\mathrm{Log}[2]\,(-2+2^{r}+\mathrm{TC}-2^{r}\,\mathrm{Log}[2])^{\frac{3-\mathrm{Log}[4]}{-1+\mathrm{Log}[2]}} \tag{4}$$

As $r$ varies from zero to Infinity, the curve takes on a 'bell' shape with a peak at position 'rpeak' on the $r$-axis and going to zero at Infinity. This 'rpeak' depends only from the TC parameter by means of a logarithmic law :

$$\mathrm{rpeak}\ = \frac{\mathrm{Log}[2\,(-2+\mathrm{TC})]}{\mathrm{Log}[2]} \tag{5}$$

***The possible correlation with mortality curves***

Equation (4) shows a function that depends on the variable r . As already mentioned, this continuous variable is linked to the recursive variable that identifies the time steps in the evolution of the cellular automaton ArbO. Now -in the general ArbO model scheme- it would be in principle possible to associate the mTC with biological species capable of developing decision-making capacities between basic alternatives including death issues possibilities, i.e. such that they can be modelled with our Arbitrary Oscillator discussed above. But if we wish to extend this general mechanism of evolution to such real situations, we must introduce the following relationship that links the generic time step to the organism's life-time:

$$r\ =\ a\ /\ \mathrm{u} \tag{6}$$

where 'a' may represent the age of the evolving organism and 'u' a time constant characteristic of the specific organism. Both a and u will be expressed in the same unit of time. For example, we may have a and u expressed in days, weeks, months or years. An e.g. value of r equal to 20 can thus be obtained from the ratio of a=100 days and u=5 days or the ratio of a=100 years and u=5 years. This last example therefore means that -after 20 steps of time intervals 'u'- the age of 100 years will be reached. It will then be possible to associate the generic ArbO clock variable $r$ with the age $a$ of the individuals considered in the demographic Life Tables. The 'u' will be linked to the specific biological species and must be considered as a constant. These different ArbO modeled species will show a mortality trend with a peak age '*apeak*' positioned on the age a-axis. This *apeak* will then be given by the following formula coming from the eq. (5):

$$apeak = u \frac{\text{Log}[2(-2 + \text{TC})]}{\text{Log}[2]} \tag{7}$$

Our mTC will then become, replacing $r$ with $(a/\text{u})$ in the eq. (4), a function of $a$ and TC:

$$\text{mTC}(a, \text{TC}) = -4^{a/\text{u}} (-1 + \text{TC} - \text{Log}[2])^{1+\frac{1}{1-\text{Log}[2]}} (-2 + \text{Log}[2]) \text{Log}[2] \left(-2 + 2^{a/\text{u}} + \text{TC} - 2^{a/\text{u}} \text{Log}[2]\right)^{\frac{3-\text{Log}[4]}{-1+\text{Log}[2]}}$$

This mTC($a$,TC) will then represent the number of deaths due to the theoretical component expressed by the ArbO model, in the interval of amplitude u between $a$-u and $a$, with $a$ as a continuous sliding variable of age. In our case of interest, i.e. that of demographic mortality expressed in human Life Tables, we will take a value u = 5 years as a constant figure for all subsequent considerations and computations. Our mTC then will become:

$$\text{mTC}(a, \text{TC}) = -4^{a/5} (-1 + \text{TC} - \text{Log}[2])^{1+\frac{1}{1-\text{Log}[2]}} (-2 + \text{Log}[2]) \text{Log}[2] \left(-2 + 2^{a/5} + \text{TC} - 2^{a/5} \text{Log}[2]\right)^{\frac{3-\text{Log}[4]}{-1+\text{Log}[2]}} \tag{8}$$

### *The general shape of the curve*

To visualise the shape of mTC(a,TC) on the age axis, we will need to choose an arbitrary conventional value of TC, which we will take e.g. as TC=100000. Fig. 2 shows the plots of the function mTC*(a,*TC*)* (in the following called "mTC") for the e.g. case where TC = 100000. The mTC values are called ‘dx counts’ for similarity with the Life Tables mortality data. It is also shown the $a$ value (*apeak*) that leads to the curve maximum (in this case *apeak* ~ 88 years). It appears that, with reference to Fig. 1, the mTC curve the areas A and B, introduced in Section 1, are almost negligible or absent. The mTC curve shows also a slight left asymmetry around the curve peak.

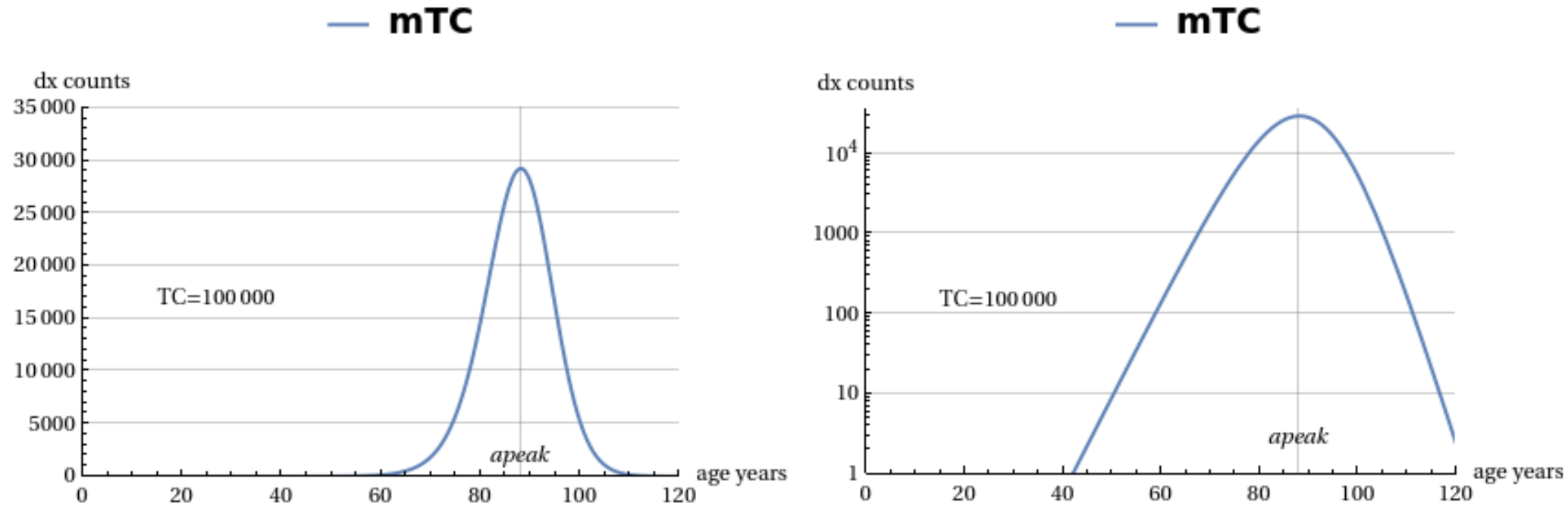

Fig. 2 - Linear and Log plot of the *m(a,*TC*)* as function of *a,* with TC=100000

### *The definite integrals*

An interesting property of mTC($a$,TC), coming from eq.’s (3) above, is also that the area under the curve itself, is proportional to the parameter TC and the continuous equations of the system S are still valid, as per the following formulae

$$\int_0^\infty \text{mTC}(a, \text{TC}) \, da = 5\,\text{TC}; \quad \int_0^\infty \text{mTC}(a, \text{TC})\, 2^{-a/u} \, da = 5; \tag{9}$$

### *The TC due to an apeak value*

The *apeak* value depends only by TC, i.e. fixed a TC value we obtain a curve with the *apeak* of eq. (7)

$$apeak = 5 \frac{\text{Log}[2(-2+\text{TC})]}{\text{Log}[2]}$$

The reverse also holds in the sense that if we deal with a mTC function and we define an *apeak* value this leads to a unique TC parameter value as per eq. (5).

$$\text{TC} = 2 + 2^{-1+\frac{\text{apeak}}{5}} \tag{10}$$

### *The mTC 'normalized' form*

The mTC function area under the curve is proportional to TC according to eq. (9). This means that e.g. with a TC value growing we will have an mTC curve expanding in size and position on age axis. This size expanding effect could be avoided by dividing the mTC by TC leading to a mTCn function:

$$\mathrm{mTCn}(a,\ \mathrm{TC}) = \left(-4^{a/5}\,(-1+\mathrm{TC}-\mathrm{Log}[2])^{1+\frac{1}{1-\mathrm{Log}[2]}}\,(-2+\mathrm{Log}[2])\,\mathrm{Log}[2]\left(-2+2^{a/5}+\mathrm{TC}-2^{a/5}\,\mathrm{Log}[2]\right)^{\frac{3-\mathrm{Log}[4]}{-1+\mathrm{Log}[2]}}\right)\Big/\mathrm{TC} \tag{11}$$

With the normalized form we can easily study the various curves shapes vs different TC values, as in Fig. 3. Here the mTCn's are plotted over nine orders of magnitude of TC , namely from 10, 100,...,10^9. We see that the *apeak* varies as expected shifting the plots to high *apeaks.* However, with exception of the initial plots, the other curves shapes look like very similar in shape and maximum height.

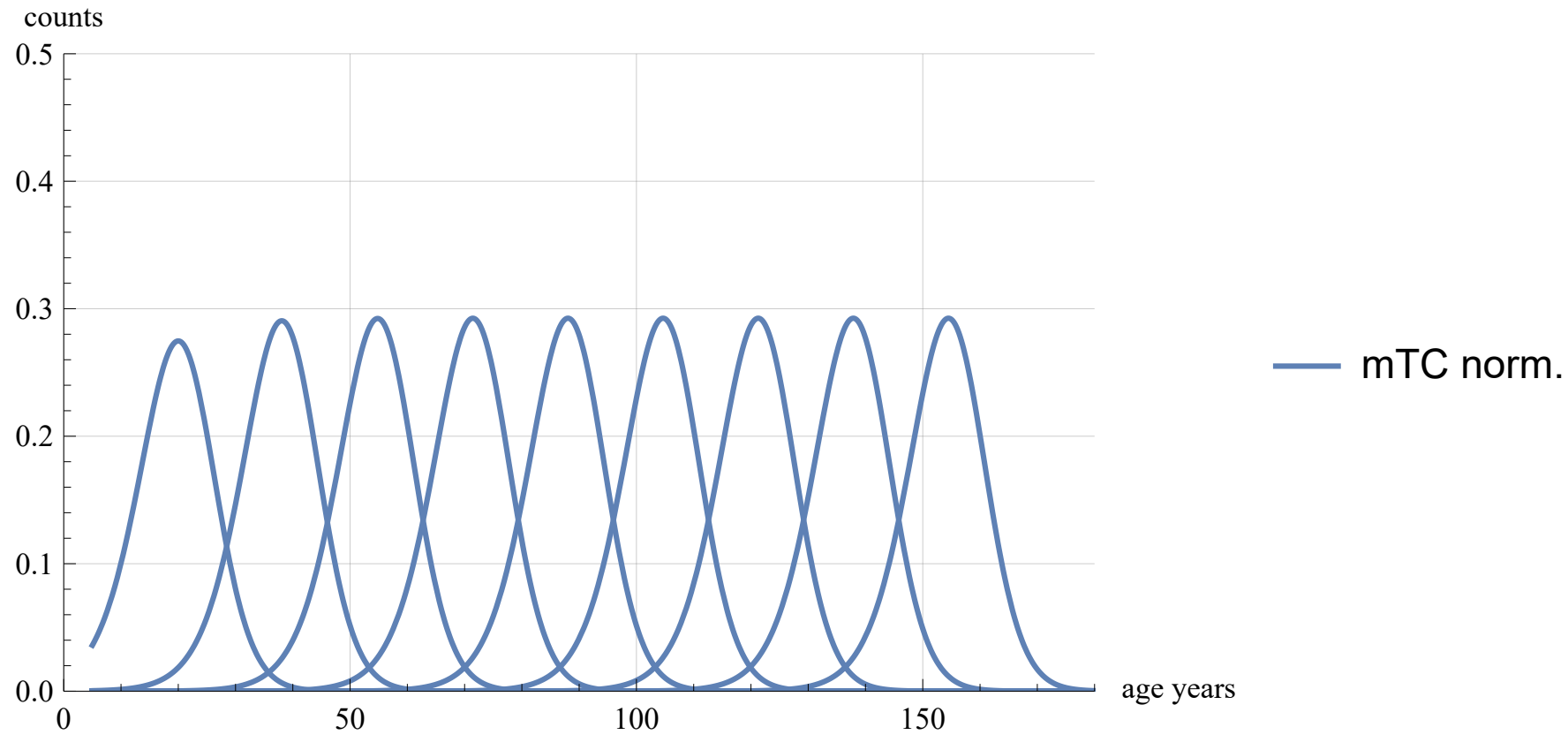

Fig.3 - Plots of normalized mTCn with TC=10^n, with n=1,2,..,9

### *The mTCnp function*

It can be useful for the further analysis to express the mTC also as function of *a* and *apeak* (=ap) by using the (10) relation. We obtain:

$$\mathrm{mTCnp}(a,\ \mathrm{ap}) = -\frac{1}{4+2^{\mathrm{ap}/5}}\,2^{2\,a/5}\left(1+2^{-1+\frac{\mathrm{ap}}{5}}-\mathrm{Log}[2]\right)^{\frac{1}{1-\mathrm{Log}[2]}}(-2+\mathrm{Log}[2])\,\mathrm{Log}[2]\left(2^{a/5}+2^{-1+\frac{\mathrm{ap}}{5}}-2^{a/5}\,\mathrm{Log}[2]\right)^{\frac{3-\mathrm{Log}[4]}{-1+\mathrm{Log}[2]}}\left(2+2^{\mathrm{ap}/5}-\mathrm{Log}[4]\right) \tag{12}$$

### *The mTCnx function and the S-system distribution*

A further step can be taken, in describing the mTC function, again by means of algebraic transformations. We formalize a mTCnx function where $x = a$-ap. In this way, we want to study the shape of the normalised mTC curve when the age variable x varies around the peak of maximum mortality. This function mTCnx becomes :

$$\begin{aligned}\mathrm{mTCnx}(x,\ \mathrm{TC}) = &-\frac{1}{\mathrm{TC}}\,2^{2\,x/5}\,(-2+\mathrm{TC})^2\,(-1+\mathrm{TC}-\mathrm{Log}[2])^{1+\frac{1}{1-\mathrm{Log}[2]}}\\ &(-2+\mathrm{Log}[2])\,\mathrm{Log}[2]^{\frac{-2+\mathrm{Log}[2]}{-1+\mathrm{Log}[2]}}\,\mathrm{Log}[4]^2\left(-\left((-2+\mathrm{TC})\left(-2^{x/5}\,\mathrm{Log}[4]+\mathrm{Log}[2]\left(-1+2^{x/5}\,\mathrm{Log}[4]\right)\right)\right)\right)^{\frac{3-\mathrm{Log}[4]}{-1+\mathrm{Log}[2]}}\end{aligned} \tag{13}$$

The normalized mTCnx function is useful for showing some aspects of the S-system distribution, particularly the near-invariance of shape and maximum height as a function of TC values, as seen in Fig. 4. Here about nine out of 10 curves are concentrated on a small range of very similar heights. This is seen in the total graph as a seemingly single curve with a maximum height of about 0.29. This is in agreement with the patterning of the curves shown in Fig. 3. Moreover, we see also a common slight asymmetry around the peaks leading to a minimal left skewness (also shown in the right graph of Fig. 2). This feature is not inconsistent with the greater left lateral asymmetry of current demographic curves.

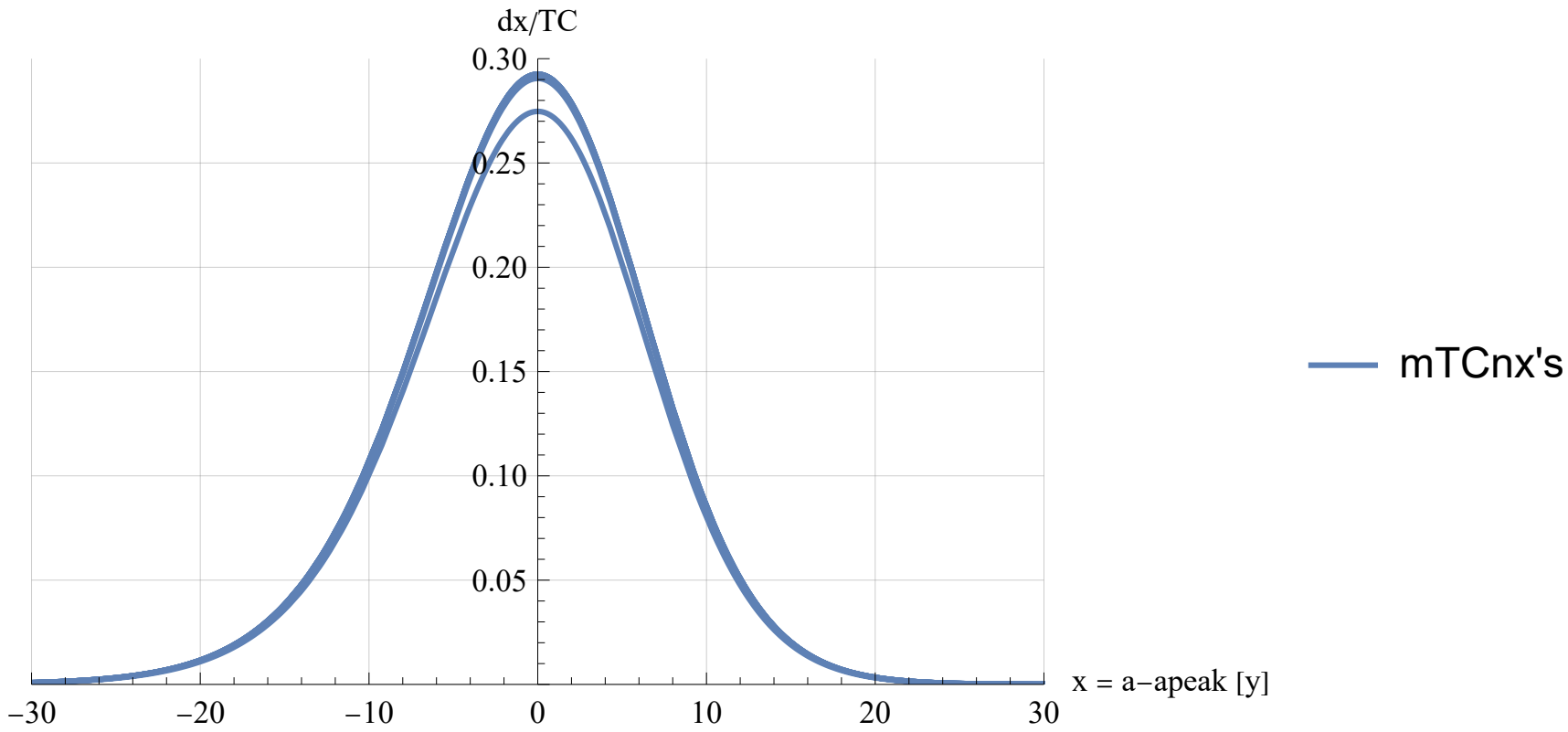

Fig.4 - Plots of normalized mTCnx with TC=10^n, with n=1,2,..,9

***Two absolute limits for the mTC***

***The height limit of the theoretical mortality curve***

Considering relation (13) above, we see that the point of maximum height of the generic curve will be given by mTCnp(ap, ap). It will then be possible to study its value as ap -and consequently TC- tend to infinity. It turns out from the calculations that:

$$\text{Limit}(\text{mTCnp}(\text{ap}, \text{ap}), \text{ap} \to \textit{Infinity}) = -4^{\frac{\text{Log}\left[\frac{\text{Log}[4]}{\text{Log}[2]}\right]}{\text{Log}[2]}} (-2 + \text{Log}[2])\, \text{Log}[2]\, (3 - \text{Log}[4])^{\frac{3-\text{Log}[4]}{-1+\text{Log}[2]}} \tag{14}$$

where the second member of eq. (14) is an absolute number whose value is 0.292541, which then in the conventional case of the Life Tables of 100000 deaths leads to a peak limit value of 29254 cases for our standard age interval of five years. This limit is asymptotic for mTCnp(ap, ap) although for practical purposes it is actually reached -in the theoretical curve case- by the age of about 60 years. Note that this limit value is a number independent from the parameter ‘u’ and therefore appears as an absolute limit for our general Life Tables max. peaks.

***The full width at half maximum limit of the theoretical mortality curve***

If, as conjectured in the present study, mortality curves are to hold up to a theoretical mTC curve as lifespan increases, then it will also be interesting to assess the ‘shape’ of these curves to see whether they approximate the theoretical shape of the mTC. This evaluation will have to be done on a measurable basis and to this end we resort to the curve FWHM (full width at half maximum). To obtain a computable FWHM, the following equation must be solved in the unknown *a*:

$$\text{mTCnp}(a, \text{ap}) = \text{mTCnp}(\text{ap}, \text{ap})/2 \tag{15}$$

In our case, the solutions can be expected to be two values, a1 and a2, with a1 < ap < a2 and the sought-after value of FWHM will be equal to

$$\text{FWHM} = \text{a2} - \text{a1}$$

Unfortunately, equation (15) is transcendent in nature and therefore very difficult to solve analytically. With numerical techniques, however, it is possible to provide quite significant results as shown in Table 1 below.

| Age peak year | 10^5 * mTCnp(ap,ap) | a1 [y] | a2 [y] | FWHM [y] |
|---|---|---|---|---|
| 15 | 26 718.1 | 6.85785 | 22.4807 | 15.6228 |
| 30 | 28 675.4 | 21.8579 | 37.4807 | 15.6228 |
| 45 | 29 175.8 | 36.8579 | 52.4807 | 15.6228 |
| 60 | 29 244.2 | 51.8579 | 67.4807 | 15.6228 |
| 75 | 29 252.9 | 66.8579 | 82.4807 | 15.6228 |
| 90 | 29 253.9 | 81.8579 | 97.4807 | 15.6228 |
| 105 | 29 254.1 | 96.8579 | 112.481 | 15.6228 |
| 120 | 29 254.1 | 111.858 | 127.481 | 15.6228 |
| 135 | 29 254.1 | 126.858 | 142.481 | 15.6228 |

Tab. 1 Reference limit FWHM computation

It can be seen from Tab. 1 that the FWHM appears invariant over the entire range of ‘ap’ values and that therefore the reference value for the FWHM to be set for the actual curves is approx. 15.6 years. The second column shows the max. height of the mTCnp and the asymptotic convergence to the max. value above defined. It should be noted, however, that while the height limit of normalised mTC is independent of the choice of species parameter u, the FWHM depends on the specific choice of u and thus can be considered constant if u is assumed to be constant (as in our case).

# 3. Comparison between the real demographic data and the mTC function

***The comparison method***

In order to check our conjecture, depicted in the Introduction, we need to compare the real Life Tables mortality data with the possible convergence with our mTC function. To this task we will use the following steps:
- for any real data set (country, survey year) interpolate the data and find the maximum *apeak* (*)
- also find the TC corresponding to this peak (eq. (10))
- by scaling, plot an mTC curve with the TC and *apeak* found and with the same maximum as the real data curve
- Also plot a "difference curve" (D-curve) obtained by subtracting the above scaled mTC from the real curve
- measure the FWHM of the interpolated real curve and the area of the D-curve and see the trend with respect to the survey years

To clarify this method, Figs. 5 and 6 are useful. In Fig. 5, the dotted curve represents actual demographic data appropriately interpolated with a continuous curve. The continuous curve below the dotted curve is our mTC function scaled to match the same peak. In Fig. 6, we have the D-curve with two vertical lines showing the location of the Fig . 5 peak and also the location of a secondary peak on the left.

---

(*) The constraint assumed for the continuous curve implies a possible shift (about 2.5 years) to the right of the simulated curve with respect to the real curve. This is not critical for our analysis, as we are looking for structural effects (not influenced by an offset component) and not for an absolute age effect.

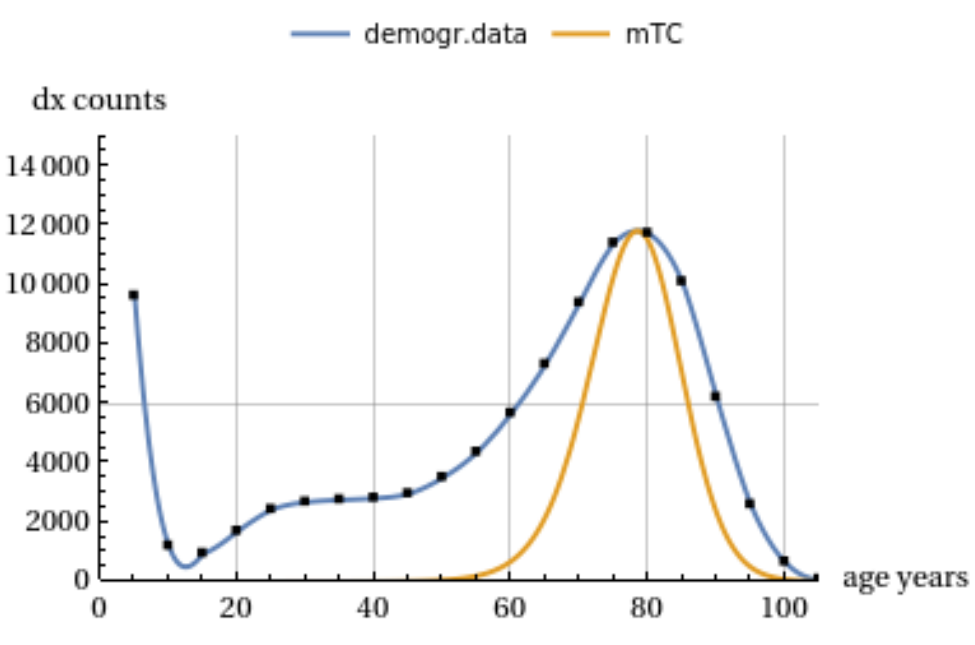

Fig.5 - Plots of interpol. demogr. data and a compliant mTC

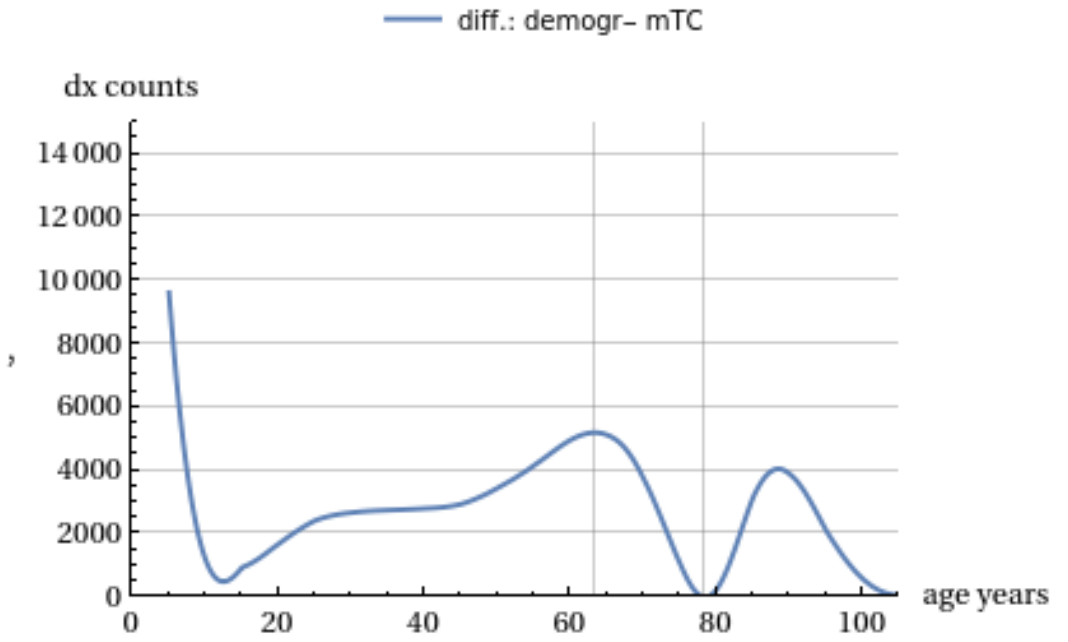

Fig.6- Plot of difference between demogr. data curve and mTC curve

***The selection of demographic data***

A large amount of mortality demographic data is available for the various countries and year of collection. We here chose to use both Italy and USA data. The ISTAT data for Italy are available at Ref. [4]. These data span a 45-years period from 1974 to 2019 (having excluded years after 2019 to avoid statistical peculiar effects due to Covid 19). The USA data were processed from a database of tables made available in the works in Ref. [5] and [6]. The USA data range runs from the year 1900 to the year 2017 mainly in ten years steps, while the Italy data are available at five years intervals. For this comparison of trends over time, it is also useful to consider differences between subsets of data that may affect the total sum result. In particular, significant differences are found between male and female mortality and also between regional subsets (not included at this stage of the study). All these different data sets were subjected to comparison with theoretical predictions of the mTC function using the method described above. The data sets used for this study are given in tables collected in the Appendix. In the following we provide some of the graphic evidences obtained and the relevant numerical tables for the main significant magnitudes. Even if total unisex data have been processed and available, we provide hereunder the above said comparison results limited for respectively USA male and female mortality and Italy male and female mortality. The graphs hereunder are presented with the format of Fig. 5 and Fig. 6. The individual discrete points correspond to the demographic data (available in the Appendix), while the curve joining them is a curve interpolated by a well-known and widely used mathematical computer application. The figures have been minified to give a quick look overview of the evolution of the shape of the curves. The main parameters (peak ages in years unit and areas in dx counts) that can be derived from the data are summarized in the table at the end of each group of figures. These tables highlight interesting data for an evaluation of our conjecture. In particular, the FWHM (years) figure can serve as a comparison with the theoretical FWHM value of an mTC curve (about 15.6 years). A measure of the total areas and individual left and right lobes of the D-curve can also show whether these curves tend to decrease with improvement in lifespan and thus whether there is convergence toward our mTC. The last column of the tables identifies a total "error" value between the possible compatible mTC and the actual curve by summing the absolute values of the two lobes and dividing them by the total area of the demographic curve. In case of perfect coincidence between mTC and real curve, this error should be zero. Note also that for data where infant mortality is very high, the TA deviates slightly from the canonical value of 100000 cases. This is due to interpolation in the initial interval that considers about 50% of the actual value.

***The USA male data comparison***

## - The over-imposed curves graphs

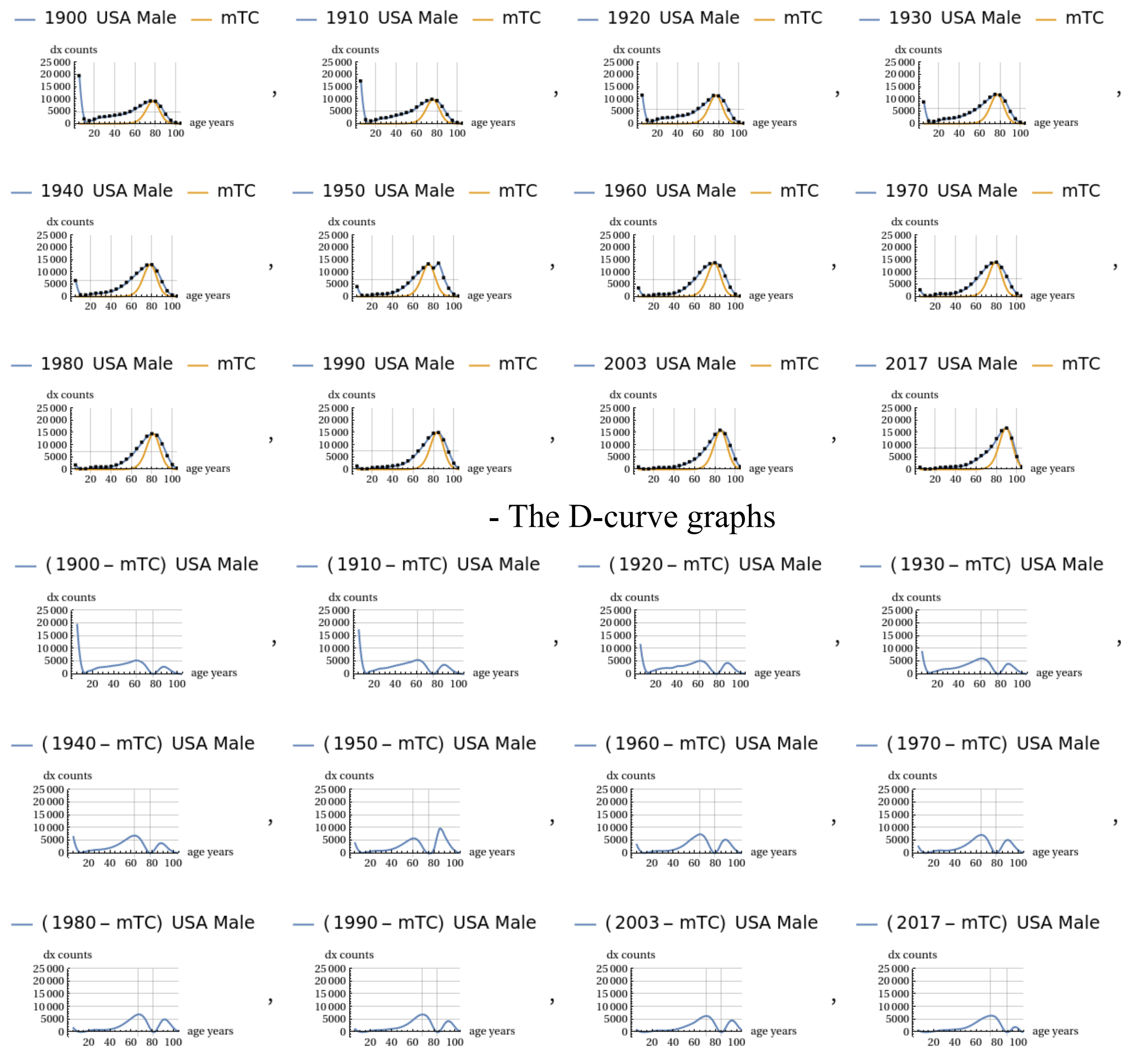

## - USA Male summary data table

| Year | Sex | 2nd Peak left | Peak | Dist. btw Peaks | FWHM | TC Param. | Area Tot.(AT) | Area Diff. | Area Diff.left(AL) | Area Diff. Right(AR) | (\|AL\|+\|AR\|)/AT |
|---|---|---|---|---|---|---|---|---|---|---|---|
| 1900 | Male | 61. | 77.1 | 16.1 | 34.6 | 21 974 | 87 438 | 55 856 | 48 955 | 6901 | 63.88% |
| 1910 | Male | 60.8 | 75.6 | 14.8 | 34.2 | 17 809 | 88 802 | 55 241 | 46 103 | 9138 | 62.21% |
| 1920 | Male | 61.8 | 77.1 | 15.2 | 30.5 | 21 776 | 92 621 | 53 265 | 42 436 | 10 829 | 57.51% |
| 1930 | Male | 61.6 | 76.9 | 15.3 | 32.2 | 21 277 | 94 386 | 53 667 | 43 668 | 9998 | 56.86% |
| 1940 | Male | 62.8 | 78. | 15.1 | 31.8 | 24 743 | 95 837 | 51 189 | 41 635 | 9555 | 53.41% |
| 1950 | Male | 60.6 | 75. | 14.4 | 33.1 | 16 357 | 97 392 | 52 295 | 30 208 | 22 088 | 53.7% |
| 1960 | Male | 65. | 78.9 | 13.9 | 32.2 | 28 129 | 97 729 | 50 953 | 37 962 | 12 991 | 52.14% |
| 1970 | Male | 64.6 | 78.6 | 14. | 31.6 | 27 038 | 98 205 | 50 484 | 37 417 | 13 066 | 51.41% |
| 1980 | Male | 67.1 | 81. | 14. | 30.3 | 37 744 | 98 714 | 49 161 | 36 601 | 12 561 | 49.8% |
| 1990 | Male | 68.9 | 83.2 | 14.3 | 29.1 | 51 167 | 98 890 | 47 252 | 37 309 | 9943 | 47.78% |
| 2003 | Male | 71.1 | 85.3 | 14.2 | 27.2 | 68 110 | 98 928 | 44 689 | 34 403 | 10 286 | 45.17% |
| 2017 | Male | 73.8 | 89. | 15.1 | 24.5 | 113 787 | 98 960 | 41 805 | 38 673 | 3132 | 42.24% |

*The USA female data comparison*

## - The over-imposed curves graphs

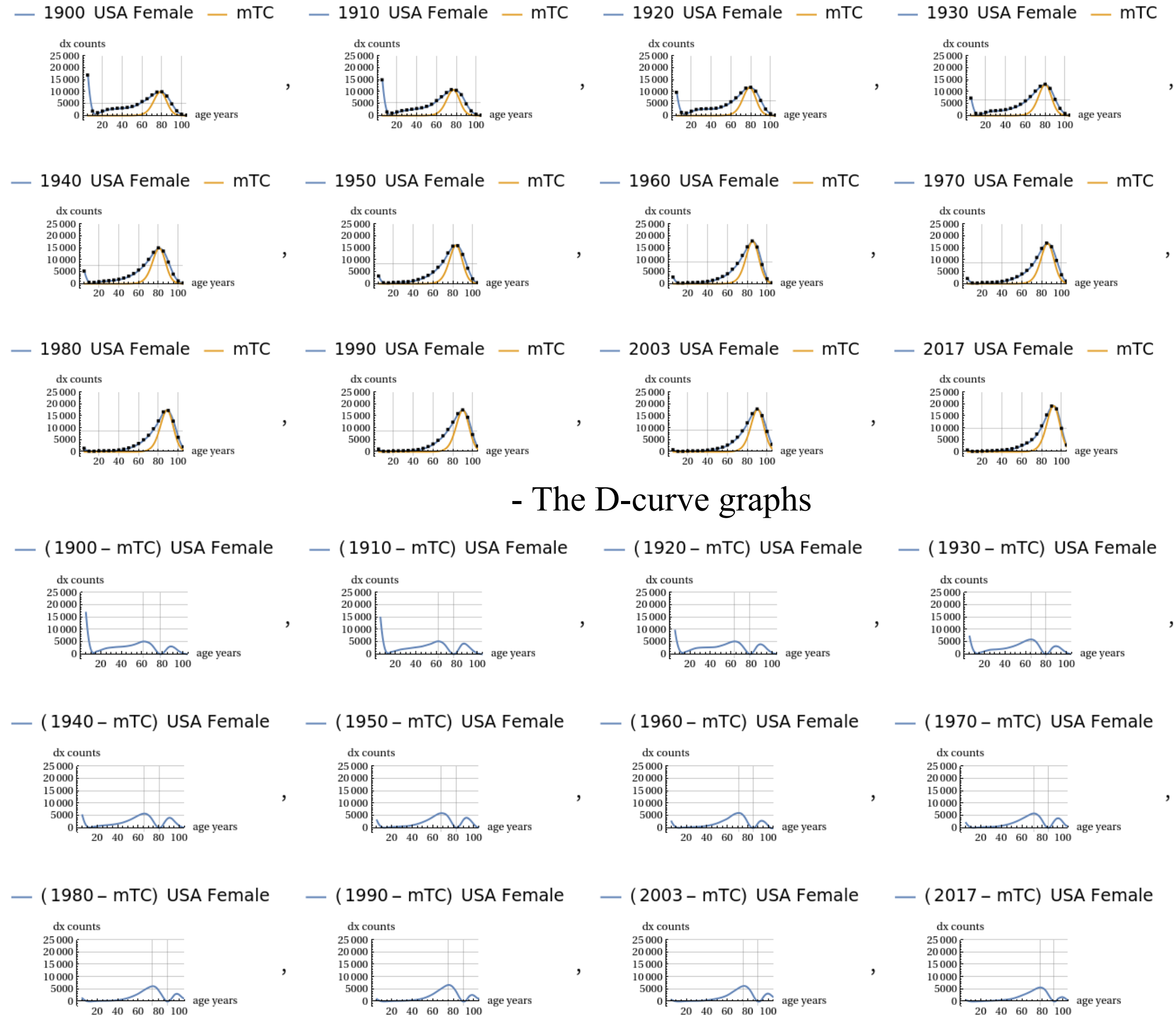

## - USA female summary data table

<table>
<tr><th>Year</th><th>Sex</th><th>2nd Peak left</th><th>Peak</th><th>Dist. btw Peaks</th><th>FWHM</th><th>TC Param.</th><th>Area Tot.(AT)</th><th>Area Diff.</th><th>Area Diff.left(AL)</th><th>Area Diff. Right(AR)</th><th>(|AL|+|AR|)/AT</th></tr>
<tr><td>1900</td><td>Female</td><td>61.7</td><td>78.1</td><td>16.4</td><td>32.6</td><td>25 052</td><td>89 123</td><td>54 965</td><td>47 217</td><td>7748</td><td>61.67%</td></tr>
<tr><td>1910</td><td>Female</td><td>61.6</td><td>76.6</td><td>15.</td><td>31.4</td><td>20 362</td><td>90 365</td><td>53 325</td><td>42 601</td><td>10 724</td><td>59.01%</td></tr>
<tr><td>1920</td><td>Female</td><td>63.3</td><td>78.4</td><td>15.1</td><td>29.5</td><td>26 381</td><td>93 820</td><td>53 405</td><td>43 345</td><td>10 060</td><td>56.92%</td></tr>
<tr><td>1930</td><td>Female</td><td>65.3</td><td>79.6</td><td>14.3</td><td>28.7</td><td>30 875</td><td>95 325</td><td>50 787</td><td>42 622</td><td>8165</td><td>53.28%</td></tr>
<tr><td>1940</td><td>Female</td><td>65.7</td><td>80.4</td><td>14.7</td><td>26.6</td><td>34 742</td><td>96 565</td><td>45 603</td><td>35 943</td><td>9660</td><td>47.23%</td></tr>
<tr><td>1950</td><td>Female</td><td>68.3</td><td>82.9</td><td>14.6</td><td>25.7</td><td>48 760</td><td>97 798</td><td>42 663</td><td>33 523</td><td>9141</td><td>43.62%</td></tr>
<tr><td>1960</td><td>Female</td><td>70.9</td><td>85.</td><td>14.1</td><td>23.5</td><td>65 538</td><td>98 000</td><td>37 277</td><td>31 941</td><td>5336</td><td>38.04%</td></tr>
<tr><td>1970</td><td>Female</td><td>71.4</td><td>85.6</td><td>14.1</td><td>24.6</td><td>70 986</td><td>98 133</td><td>40 376</td><td>31 974</td><td>8402</td><td>41.14%</td></tr>
<tr><td>1980</td><td>Female</td><td>74.</td><td>88.3</td><td>14.3</td><td>24.2</td><td>103 346</td><td>98 181</td><td>38 835</td><td>32 570</td><td>6265</td><td>39.55%</td></tr>
<tr><td>1990</td><td>Female</td><td>75.2</td><td>89.6</td><td>14.4</td><td>24.8</td><td>124 863</td><td>98 267</td><td>39 179</td><td>34 225</td><td>4954</td><td>39.87%</td></tr>
<tr><td>2003</td><td>Female</td><td>76.</td><td>90.</td><td>14.</td><td>24.2</td><td>131 069</td><td>98 257</td><td>38 090</td><td>32 076</td><td>6014</td><td>38.77%</td></tr>
<tr><td>2017</td><td>Female</td><td>77.6</td><td>91.5</td><td>14.</td><td>21.5</td><td>161 934</td><td>98 454</td><td>33 789</td><td>30 723</td><td>3066</td><td>34.32%</td></tr>
</table>

***The Italy male data comparison***

## - The over-imposed curves graphs

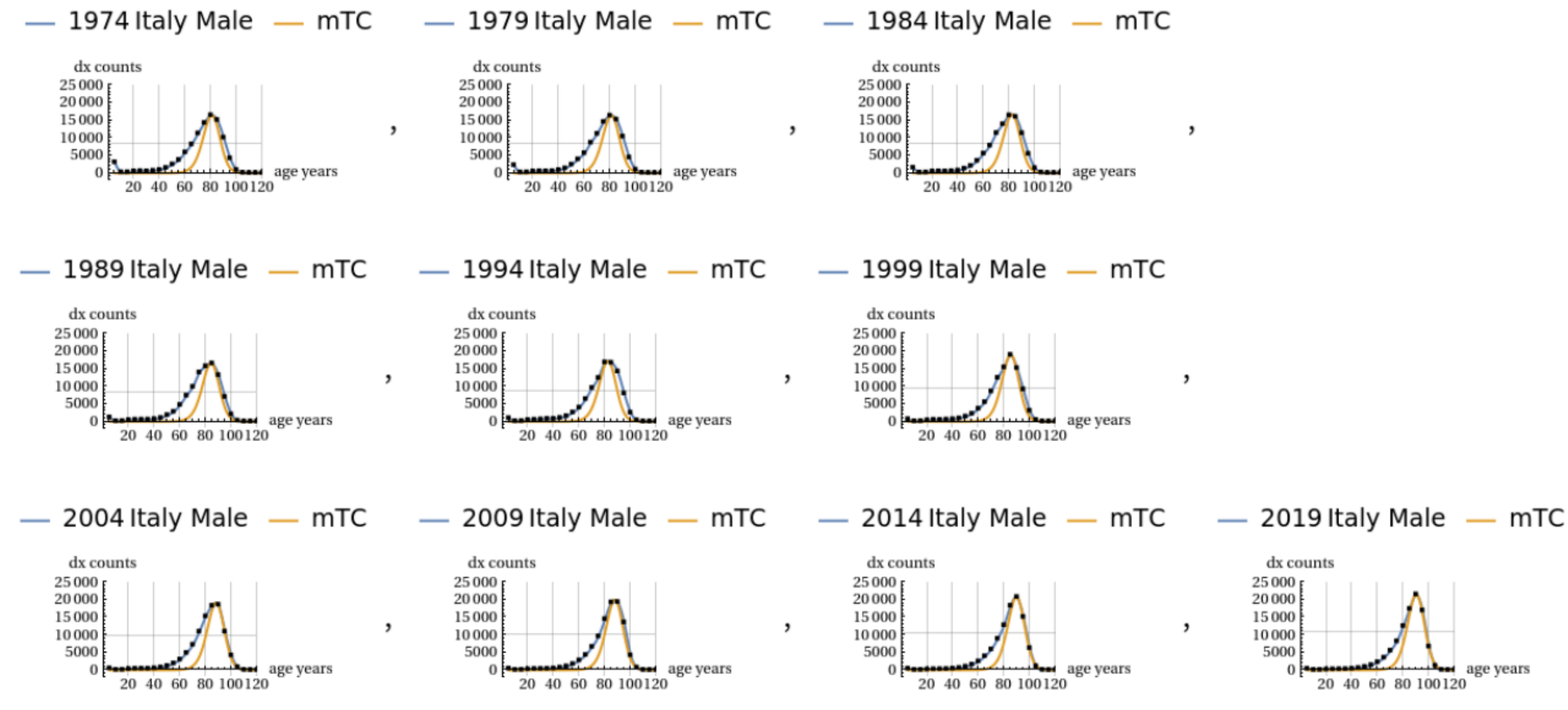

## - The D-curve graphs

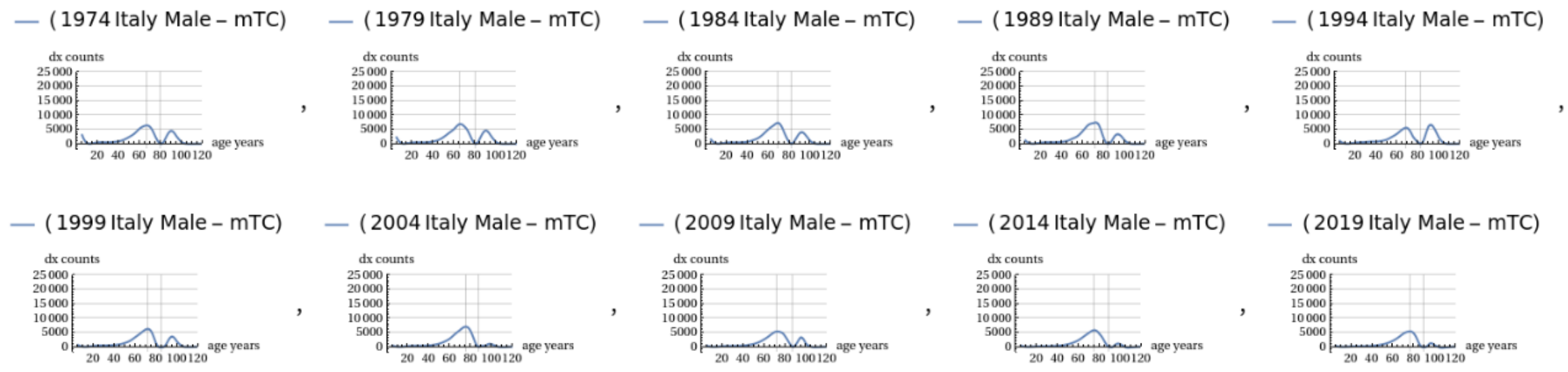

## - Italy male summary data table

| Year | Sex | 2nd Peak left | Peak | Dist. btw Peaks | FWHM | TC Param. | Area Tot.(AT) | Area Diff. | Area Diff.left(AL) | Area Diff. Right(AR) | (\|AL\|+\|AR\|)/AT |
|---|---|---|---|---|---|---|---|---|---|---|---|
| 1974 | Male | 66.9 | 80.6 | 13.7 | 26.3 | 35 761 | 98 022 | 41 864 | 31 912 | 9953 | 42.71% |
| 1979 | Male | 65.9 | 80.8 | 14.9 | 27.5 | 36 676 | 98 559 | 42 768 | 32 554 | 10 214 | 43.39% |
| 1984 | Male | 69.1 | 82. | 12.9 | 26.7 | 43 347 | 99 048 | 42 245 | 33 194 | 9052 | 42.65% |
| 1989 | Male | 72.2 | 84.1 | 11.9 | 26.9 | 57 540 | 99 265 | 42 636 | 35 618 | 7018 | 42.95% |
| 1994 | Male | 68.5 | 82.2 | 13.7 | 26. | 44 624 | 99 406 | 40 700 | 25 628 | 15 072 | 40.94% |
| 1999 | Male | 71.8 | 85. | 13.2 | 23.5 | 65 538 | 99 590 | 34 861 | 28 044 | 6817 | 35% |
| 2004 | Male | 75.4 | 88. | 12.5 | 22.6 | 98 807 | 99 724 | 34 495 | 32 610 | 1885 | 34.59% |
| 2009 | Male | 72.7 | 87.6 | 15. | 21.4 | 94 551 | 99 761 | 31 670 | 26 616 | 5054 | 31.75% |
| 2014 | Male | 75. | 89.5 | 14.5 | 20.4 | 122 737 | 99 788 | 28 797 | 27 580 | 1217 | 28.86% |
| 2019 | Male | 76.3 | 90. | 13.7 | 19.9 | 130 945 | 99 810 | 26 463 | 25 342 | 1120 | 26.51% |

*The Italy female data comparison*

## - The over-imposed curves graphs

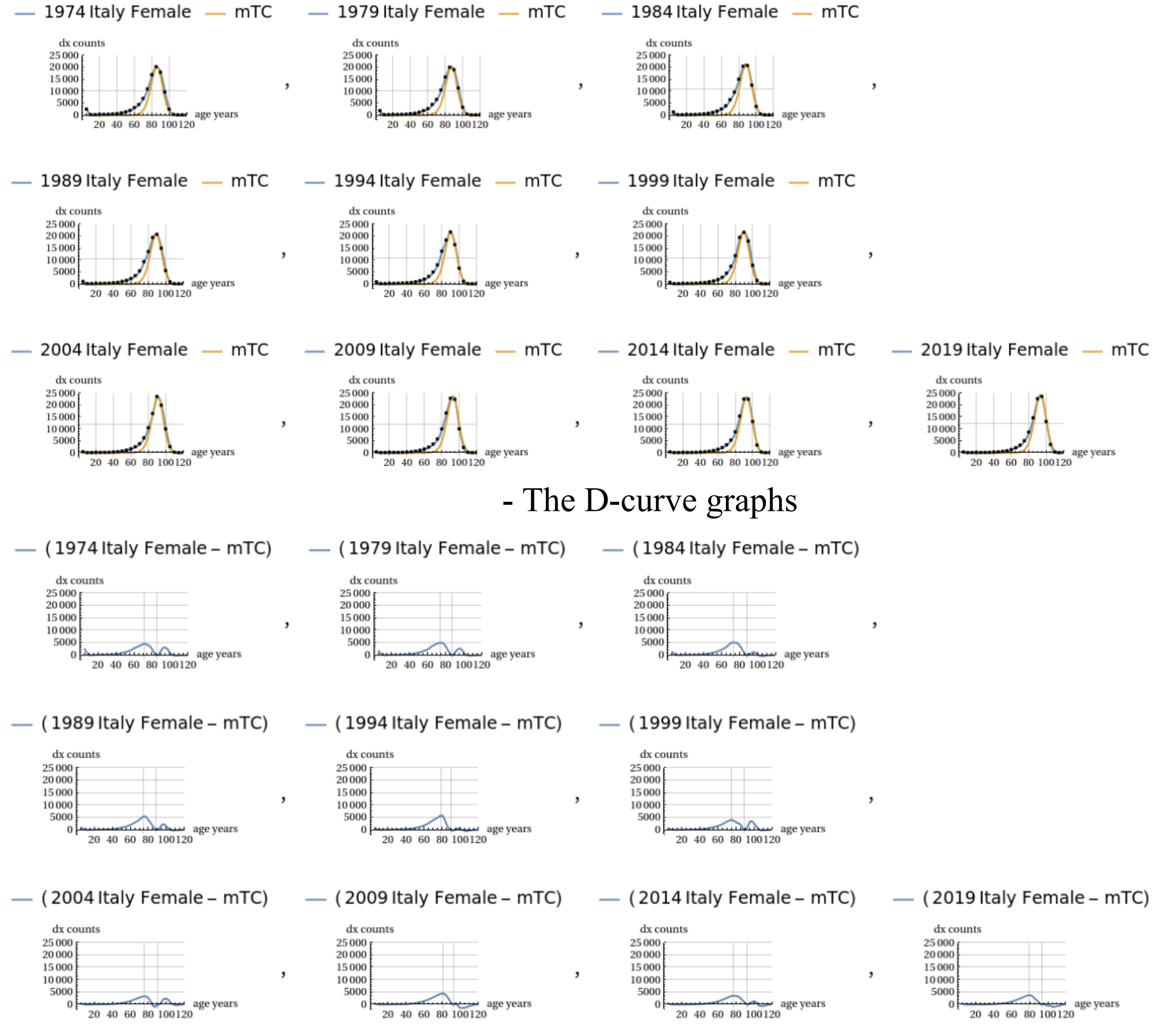

## - Italy female summary data table

| Year | Sex | 2nd Peak left | Peak | Dist. btw Peaks | FWHM | TC Param. | Area Tot.(AT) | Area Diff. | Area Diff.left(AL) | Area Diff. Right(AR) | (\|AL\|+\|AR\|)/AT |
|---|---|---|---|---|---|---|---|---|---|---|---|
| 1974 | Female | 71.5 | 85.5 | 14. | 20.5 | 69 960 | 98 428 | 29 490 | 23 848 | 5642 | 29.96% |
| 1979 | Female | 73.7 | 86.7 | 13. | 20.9 | 82 646 | 98 872 | 29 727 | 25 271 | 4457 | 30.07% |
| 1984 | Female | 73. | 87.8 | 14.8 | 19.8 | 96 897 | 99 230 | 26 291 | 25 133 | 1157 | 26.49% |
| 1989 | Female | 75. | 88.6 | 13.6 | 20.8 | 108 577 | 99 406 | 27 955 | 24 617 | 3338 | 28.12% |
| 1994 | Female | 78.6 | 90. | 11.4 | 20.1 | 131 073 | 99 523 | 25 487 | 25 664 | -177 | 25.96% |
| 1999 | Female | 74.9 | 89.4 | 14.5 | 19.4 | 119 992 | 99 658 | 25 920 | 20 161 | 5758 | 26.01% |
| 2004 | Female | 75.6 | 90.5 | 14.9 | 17.8 | 139 670 | 99 811 | 19 369 | 15 020 | 4349 | 19.41% |
| 2009 | Female | 80.8 | 92.4 | 11.6 | 17.7 | 182 288 | 99 844 | 18 963 | 21 925 | -2962 | 24.93% |
| 2014 | Female | 77. | 92.5 | 15.5 | 18.1 | 185 706 | 99 888 | 19 919 | 18 865 | 1054 | 19.94% |
| 2019 | Female | 78.9 | 93.1 | 14.3 | 17.3 | 201 847 | 99 900 | 17 120 | 18 910 | -1790 | 20.72% |

# 4. Discussion of the results

From the graphs and tables shown above for the two case studies (U.S. and Italy), there is a general trend of shrinking areas A and B and also evident is the shrinking FWHM of area C. This occurs in the face of an increase in life span. The shape of area C also approximates the theoretical shape of the mTC function as lifespan improves (improvement measurable by the shift to the right on the age axis of the mortality peak in the succession of different survey years). The process of convergence toward the mTC curve takes place not only with the disappearance of areas A and B (not really present in the mTC model) but also with the gradual narrowing (to the point of disappearing or becoming negative in some cases) of the right-hand lobe of the D-curve. Another remarkable aspect is the difference in parameters (peak, FWHM) between males and females in both case studies. In the case of italian females, the tendency to converge toward a mTC-type curve is more pronounced. The FWHM of 17.3 years, for year 2019 females, is very close to the theoretical figure of 15.6 years. The area "error" in last table column marking the difference between the actual and theoretical curves areas is also reduced to about 20% in the best case. An interesting aspect is also the relative invariance of the distance in years (around 14 years) between the main peak and the secondary peak to the left of the D-curve. This is even though the area of the left lobe tends to decrease. This invariance holds for the two study cases (USA and Italy). These findings are presented in various graphs that collect the main parameters on a single time scale from the year 1900 to 2019. In Fig. 7 and 8 the FWHM datum is given together with the theoretical FWHM level for a mTC function. The downward trend in FWHM compared to demographic surveys is present in both the U.S. and Italian data, although the absolute values of FWHM are generally higher for the United States. This can be explained by considering that the U.S. population is about a factor of five larger than that of Italy, which probably leads to a statistical dispersion between the mortality curves of the two case studies.

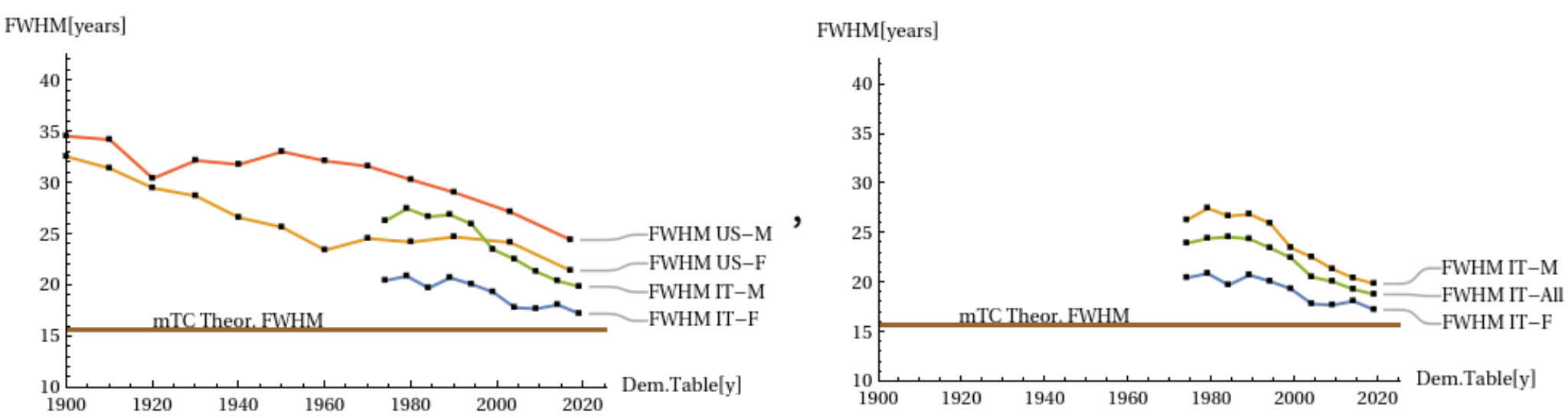

Fig.7 - FWHM USA & Italy subsets vs demogr. table year Fig. 8 - Italy FWHM showing the sex split plus total datum

In Fig. 9 the area "error" amount trend (last tables column) is shown .

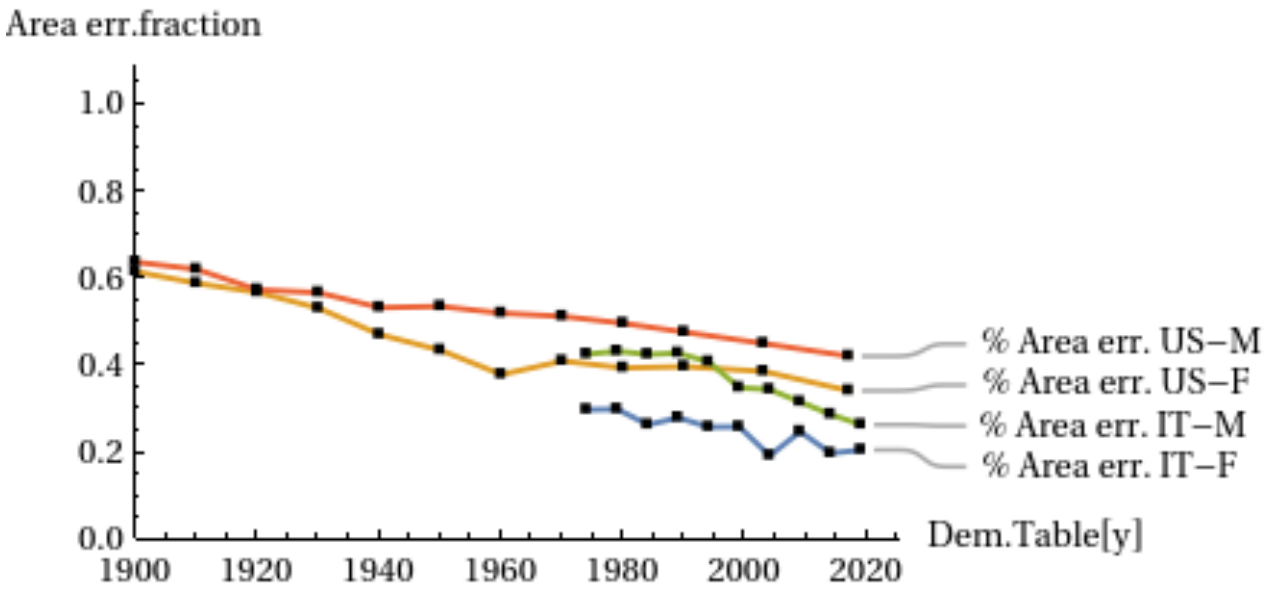

Fig.9 - Area “error” for USA & Italy subsets vs demogr. table year

Fig. 10 shows the improvement in the maximum mortality peak vs time. This is in particular evident for females in both study cases. The mix effect on "*apeak*" is also presented for Italy in Fig. 11. The overlap of the US and Italy data in the common time period is significant.

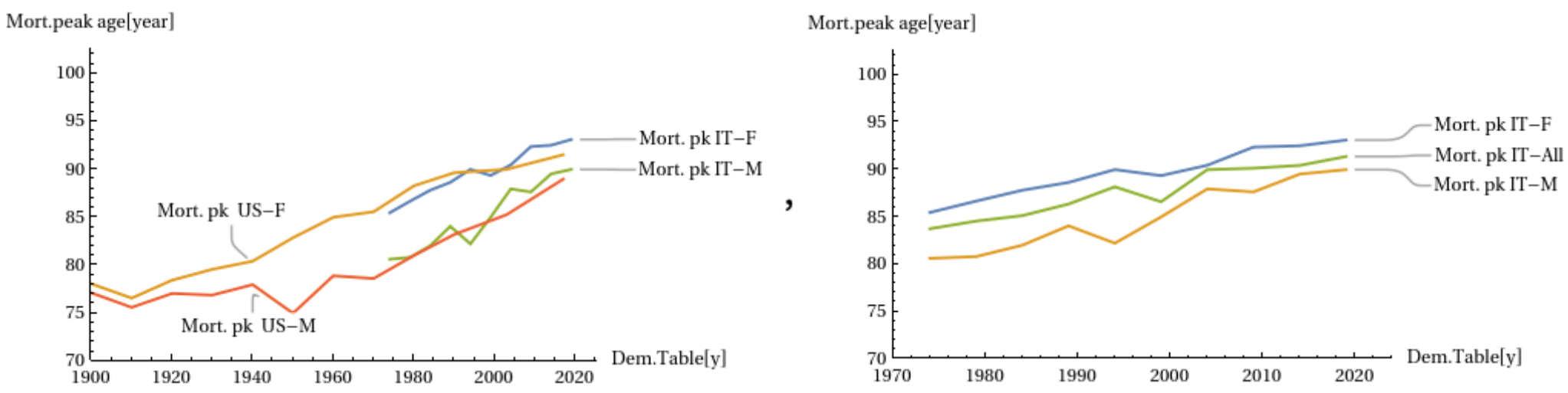

Fig. 10 - *apeak* for USA & Italy subsets vs Dem. table year

Fig. 11 - Italy *apeak* showing the sex split plus total datum

Fig. 12 shows in Log scale the evolution of the curves TC equivalent parameter, confirming the overlap of USA and Italy TC data for the common time interval.

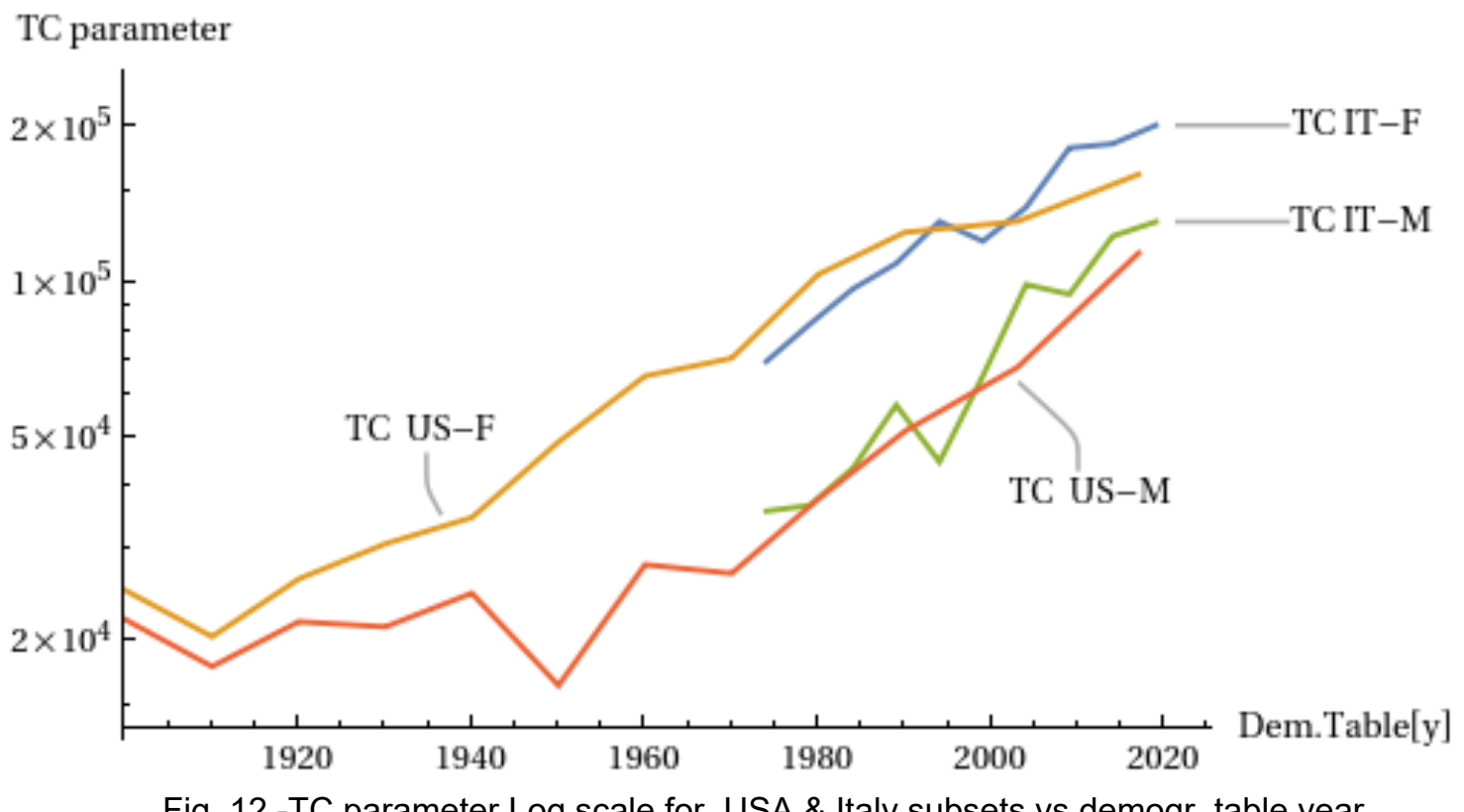

Fig. 12 -TC parameter Log scale for USA & Italy subsets vs demogr. table year

## 5. An attempt to forecast demographic mortality for the years to come

Considering the above data, there is an almost linear trend in the growth of the *apeak* value (Fig. 10). This justifies the assumption of a continuous growth in lifespan. This also allows us to try a forecast for the future evolution of demographic mortality. This can be done by using the mTCnp version of the formula (8) and assuming a future peak change in continuity with the rate of change recorded in the last years of population surveys. The result of this simulation is shown in Fig. 13 for the case of Italian females. The figure shows the latest demographic curves (2009 and 2019) along with the theoretical projections of the mTCnp function -multiplied by 100000 for commonalty with standard sampled data- when it assumes a constant shift rate for '*apeak*' values for the years 2029, 2039, 2049. These peak values are calculated with the same peak growth rate recorded for last years survey. It is interesting to note that, as predicted in Section 2, the theoretical mortality curves, while shifting on the age axis, maintain the same shape and height at 29254 deaths (over 100000 total) on the peak mortality. Even this feature that defines an invariant shape and a constant upper limit number of peak mortality may be useful for validating -or not validating- the conjecture in the future.

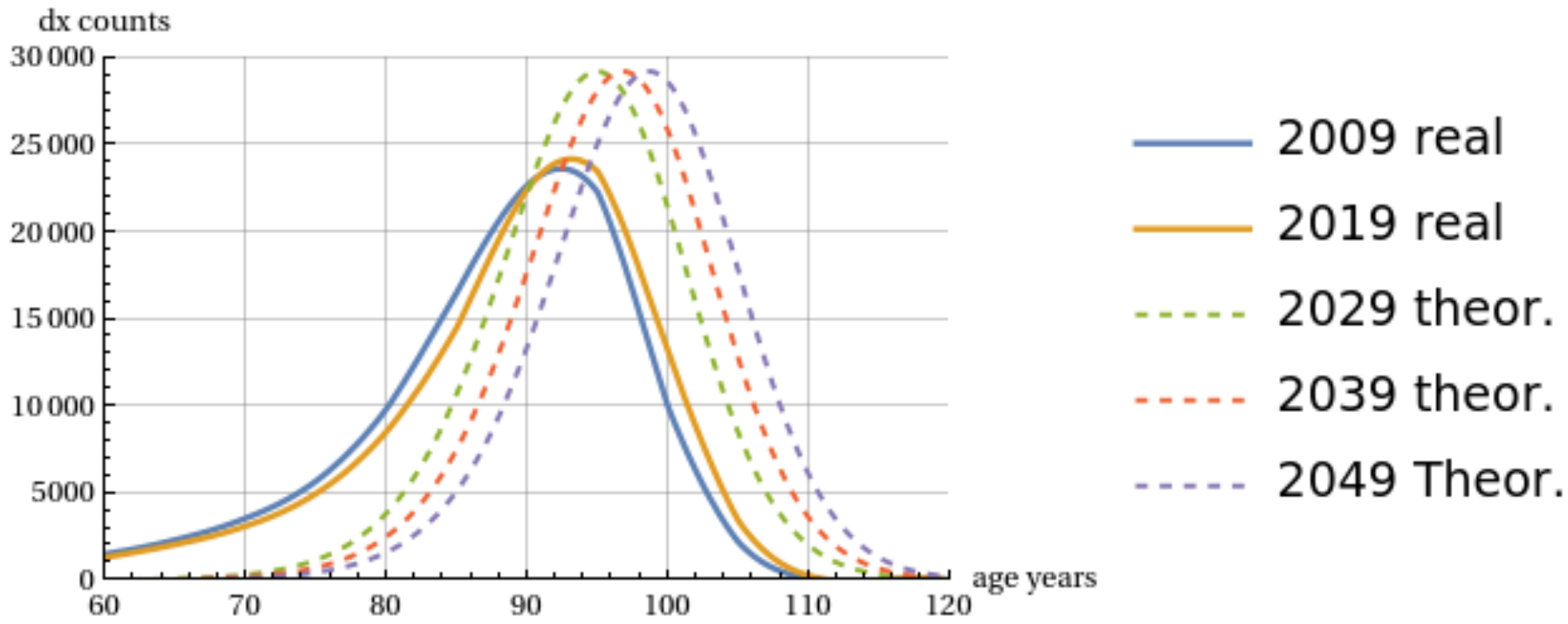

Fig. 13 - Possible future evolution of demographic mortality according to the conjecture (Italy female case)

Of course, in this hypothetical prediction the demographic curve must also confirm the trend of narrowing left skewness, i.e. we must hypothesize an almost disappearance of A and B areas in Life Tables. If this last trend were less fast, than the above forecast might be different. In other words, the future curve for 2029 may catch up with the theoretical curve by evolving in the 2019-2029 interval or come close to it without concurring with it and leaving the final overlap for later years. This will depend on the speed of reduction of the left skewness. We can summarise here the features that the conjecture would imply about future mortality curve projections, in particular involving the theoretical limits described in Section 2:

- the peak mortality will never exceed 29254 cases in a 5-year sampling interval
- the FWHM of the curves cannot decline indefinitely but will have a lower limit of 15.6 years with a constant u=5 years
- the slope of the descent curve to the right of the peak will be constant regardless of lifespan improvement

A detailed study on the actual practical existence of these theoretical limits in relation to real population curves was made by the author in Ref. [7].

# 6. Conclusions, validity of the conjecture and aging theories

The present work is based on a model (the Arbitrary Oscillator - ArbO, Ref. [1]) that generates events that evolve over time and may also include instances of “death.” To find the most probable distribution of such events, the Fermi statistic has been used, obtaining a recursive computational algorithm that provides the solution as dependent on a single parameter TC. By means of scaling, the mortality predictions generated by the algorithm can be compared with actual demographic mortality data. This comparison shows similarities of the theoretical curves generated by the algorithm with the shape of the real data. From the Sec. 3 results, the convergence of the real curves toward the theoretical mTC curve as the apeak value increases seems to be realistic. Therefore, the conjecture of the trend convergence of the real mortality data toward a theoretical S-System type curve in the presence of increasing life span seems reasonable. However, there persists a component of the D curve, namely the left lobe area, which - while trending downward as expected - does not decrease with the same speed as the right lobe area in the time sequence. This fact may suggest a complex structure of components underlying the total real curve. In fact, consider, for example, Fig. 11, where a mix effect of the sum of male and female mortality on total mortality is observed, leading to a diversification of the total mortality peak with a shift to the left from the highest peak in the female group. Similarly, there could be diversified components of the theoretical mTC curves that add up generating the leftward bias of the actual curves. In this hypothesis, the trend curve of the theoretical mTC would not be unique, but given by a group or "spectrum" of components of the theoretical mTC's. These sub-components could add up to produce the actual curve. This further analysis can be a research ground for further development and verification of the conjecture. See in this connection also Ref. [8] where a tentative explanation of these sub-components is given. As a final comment it should be noted that the model to which the conjecture refers is not the result of application of predictive "filters" on data sequences or of exercises in fitting statistical mortality data or an “a-priori” assumption of a mortality probability distribution, but an extension of a theoretical mathematical model (the ArbO model) when applied to demographic mortality data. The assumed issue of lifespan extension is obviously linked to the issue of aging modelling. Some theories refer to the reliability model Ref. [9-10] and failure accumulation Ref. [11]. Other theories explain aging as a statistical effect Ref. [12-13]. Some other theories predict both the possibility of an indefinite extension of lifespan Ref. [14-15] and -on the contrary- the implausibility of this hypothesis Ref. [16]. Finally, there is also the hypothesis that aging is an effect inherent in evolution Ref. [17-18]. Our conjecture does not provide evidence to determine whether increased longevity is an effect that can continue or whether it is destined to stop. In fact, if the TC parameter is increased, the curves shift towards greater longevity. However, we do not know whether TC will continue to grow, as appears from the data collected so far, or whether it will stop in the future. Certainly, in our model, the aging of a population is something structurally imposed by the limit on the number of possible 'critical' cases, i.e., by the TC parameter. Once TC is set for a population of organisms following the ArbO model, the accumulation of critical events will inevitably approach the accumulation limit of 'fatal' events, leading to 'critical' events always deadly ended. Aging, i.e. the approaching to the end of life with advancing age, therefore appears to be an inevitable statistical fact predicted by the 'more probable' solution of the equations set of the ArbO model.

# References

Author Note on References:

*The literature on demographic mortality is enormous and covers a broad spectrum of scientific and social aspects. It is beyond the scope and capacity of this article to attempt to review and make reference with these studies. We will therefore only refer to works strictly instrumental to the contents of this work, leaving the correlation with other studies on the topic to the interest and experience of the reader. This choice can also be justified - in the author's opinion - by the fact that the approach illustrated here is original and not directly related to other works and research models.*

# Appendix

In the following pages we provide the demographic data tables at the origin of the study.

| Age interv. [y] | 1900 | 1910 | 1920 | 1930 | 1940 | 1950 | 1960 | 1970 | 1980 | 1990 | 2003 | 2017 |
|---|---|---|---|---|---|---|---|---|---|---|---|---|
| 0–4 | 19 452 | 17 282 | 11 495 | 8706 | 6376 | 3923 | 3357 | 2605 | 1667 | 1246 | 900 | 739 |
| 5–9 | 1773 | 1469 | 1321 | 948 | 570 | 351 | 268 | 244 | 173 | 127 | 80 | 62 |
| 10–14 | 1094 | 988 | 1028 | 785 | 546 | 360 | 268 | 247 | 188 | 163 | 114 | 92 |
| 15–19 | 1697 | 1469 | 1716 | 1341 | 891 | 671 | 616 | 778 | 656 | 610 | 456 | 358 |
| 20–24 | 2512 | 2117 | 2188 | 1861 | 1232 | 904 | 860 | 1086 | 955 | 805 | 689 | 678 |
| 25–29 | 2725 | 2297 | 2362 | 2013 | 1376 | 930 | 805 | 968 | 931 | 883 | 650 | 836 |
| 30–34 | 2995 | 2764 | 2376 | 2271 | 1638 | 1101 | 937 | 1075 | 929 | 1075 | 720 | 951 |
| 35–39 | 3305 | 3317 | 3082 | 2718 | 2125 | 1553 | 1317 | 1456 | 1156 | 1330 | 970 | 1088 |
| 40–44 | 3598 | 3779 | 3188 | 3475 | 2910 | 2388 | 2080 | 2172 | 1696 | 1622 | 1440 | 1293 |
| 45–49 | 4113 | 4400 | 3691 | 4364 | 4082 | 3661 | 3293 | 3299 | 2642 | 2274 | 2135 | 1798 |
| 50–54 | 4797 | 5148 | 4588 | 5537 | 5627 | 5377 | 5160 | 4931 | 4071 | 3373 | 3000 | 2740 |
| 55–59 | 6044 | 6627 | 6048 | 7072 | 7485 | 7535 | 7152 | 7181 | 5924 | 5114 | 4121 | 4021 |
| 60–64 | 7159 | 8079 | 7699 | 8755 | 9366 | 9680 | 9710 | 9640 | 8366 | 7407 | 6032 | 5506 |
| 65–69 | 8519 | 9241 | 9550 | 10 638 | 11 188 | 11 616 | 11 933 | 12 022 | 10 965 | 9864 | 8312 | 7053 |
| 70–74 | 9141 | 9810 | 11 352 | 11 798 | 12 724 | 13 194 | 13 294 | 13 499 | 13 409 | 12 722 | 11 153 | 9261 |
| 75–79 | 8992 | 9271 | 11 188 | 11 546 | 12 869 | 11 519 | 13 650 | 13 876 | 14 462 | 14 636 | 14 148 | 12 429 |
| 80–84 | 6905 | 6883 | 9208 | 9065 | 10 302 | 13 487 | 12 455 | 11 753 | 13 790 | 14 934 | 15 892 | 15 656 |
| 85–89 | 3671 | 3557 | 5393 | 4824 | 5906 | 7553 | 8236 | 8061 | 10 288 | 11 937 | 14 445 | 16 752 |
| 90–94 | 1246 | 1213 | 1971 | 1832 | 2201 | 3242 | 3639 | 3781 | 5453 | 6951 | 9633 | 12 617 |
| 95–99 | 240 | 256 | 494 | 411 | 508 | 834 | 853 | 1104 | 1856 | 2398 | 4085 | 5099 |
| >100 | 22 | 33 | 62 | 40 | 78 | 121 | 117 | 222 | 423 | 529 | 1025 | 971 |

**USA mortality 1900–2017, Male**

| Age interv. [y] | 1900 | 1910 | 1920 | 1930 | 1940 | 1950 | 1960 | 1970 | 1980 | 1990 | 2003 | 2017 |
|---|---|---|---|---|---|---|---|---|---|---|---|---|
| 0–4 | 16 881 | 14 883 | 9620 | 7211 | 5152 | 3092 | 2629 | 2045 | 1334 | 994 | 718 | 607 |
| 5–9 | 1729 | 1389 | 1194 | 781 | 446 | 256 | 198 | 171 | 122 | 95 | 65 | 52 |
| 10–14 | 1083 | 915 | 939 | 644 | 402 | 221 | 157 | 148 | 112 | 97 | 74 | 61 |
| 15–19 | 1752 | 1395 | 1691 | 1248 | 707 | 365 | 260 | 305 | 248 | 217 | 194 | 146 |
| 20–24 | 2436 | 1937 | 2421 | 1788 | 971 | 483 | 338 | 365 | 301 | 272 | 246 | 251 |
| 25–29 | 2725 | 2234 | 2672 | 1930 | 1140 | 650 | 422 | 422 | 332 | 312 | 273 | 340 |
| 30–34 | 2931 | 2528 | 2750 | 2094 | 1372 | 727 | 587 | 578 | 411 | 417 | 358 | 460 |
| 35–39 | 3056 | 2825 | 2806 | 2377 | 1718 | 1105 | 849 | 869 | 609 | 563 | 562 | 590 |
| 40–44 | 3286 | 3139 | 2953 | 2886 | 2236 | 1632 | 1295 | 1304 | 961 | 811 | 869 | 796 |
| 45–49 | 3706 | 3754 | 3502 | 3585 | 3028 | 2394 | 1938 | 1941 | 1510 | 1290 | 1267 | 1154 |
| 50–54 | 4507 | 4609 | 4353 | 4624 | 4120 | 3381 | 2876 | 2786 | 2300 | 2051 | 1808 | 1775 |
| 55–59 | 5753 | 6166 | 5661 | 6037 | 5615 | 4804 | 4021 | 3927 | 3346 | 3139 | 2694 | 2606 |
| 60–64 | 6909 | 7788 | 7312 | 7871 | 7570 | 6771 | 5968 | 5441 | 4894 | 4667 | 4159 | 3566 |
| 65–69 | 8525 | 9522 | 9385 | 10 150 | 10 074 | 9246 | 8362 | 7743 | 6800 | 6553 | 6001 | 4959 |
| 70–74 | 9727 | 10 761 | 11 397 | 12 174 | 13 024 | 12 762 | 11 706 | 10 848 | 9534 | 9235 | 8633 | 7293 |
| 75–79 | 9865 | 10 473 | 11 731 | 13 022 | 14 901 | 15 625 | 15 331 | 14 662 | 12 814 | 12 301 | 12 018 | 10 753 |
| 80–84 | 8066 | 8631 | 10 098 | 11 256 | 13 552 | 15 818 | 17 794 | 16 907 | 16 600 | 15 871 | 15 870 | 15 327 |
| 85–89 | 4757 | 4782 | 6201 | 6666 | 8928 | 12 120 | 15 213 | 15 378 | 17 194 | 17 449 | 17 803 | 19 042 |
| 90–94 | 1854 | 1828 | 2586 | 2849 | 3849 | 6341 | 7863 | 9595 | 12 716 | 14 320 | 15 027 | 17 839 |
| 95–99 | 409 | 392 | 656 | 725 | 1016 | 1909 | 1929 | 3611 | 5935 | 7095 | 8408 | 9686 |
| >100 | 43 | 49 | 72 | 82 | 179 | 298 | 264 | 954 | 1927 | 2251 | 2952 | 2697 |

**USA mortality 1900–2017, Female**

| Age interv. [y] | **1974** | **1979** | **1984** | **1989** | **1994** | **1999** | **2004** | **2009** | **2014** | **2019** |
|---|---|---|---|---|---|---|---|---|---|---|
| 0–4 | 3042 | 2219 | 1487 | 1149 | 949 | 666 | 473 | 410 | 389 | 363 |
| 5–9 | 214 | 172 | 138 | 103 | 101 | 72 | 58 | 46 | 42 | 39 |
| 10–14 | 213 | 194 | 158 | 131 | 129 | 106 | 74 | 70 | 50 | 49 |
| 15–19 | 478 | 429 | 419 | 389 | 416 | 323 | 260 | 218 | 148 | 138 |
| 20–24 | 538 | 503 | 477 | 523 | 518 | 499 | 367 | 305 | 224 | 206 |
| 25–29 | 479 | 483 | 447 | 531 | 606 | 475 | 383 | 329 | 248 | 229 |
| 30–34 | 634 | 522 | 518 | 548 | 758 | 556 | 380 | 360 | 290 | 265 |
| 35–39 | 862 | 833 | 690 | 682 | 756 | 653 | 502 | 444 | 398 | 354 |
| 40–44 | 1441 | 1320 | 1237 | 1006 | 1016 | 884 | 724 | 676 | 580 | 572 |
| 45–49 | 2424 | 2372 | 2004 | 1809 | 1509 | 1393 | 1142 | 1030 | 947 | 883 |
| 50–54 | 3686 | 3869 | 3451 | 2751 | 2553 | 2189 | 1865 | 1678 | 1526 | 1393 |
| 55–59 | 5839 | 5623 | 5488 | 4694 | 3938 | 3633 | 2961 | 2729 | 2433 | 2213 |
| 60–64 | 8109 | 8608 | 7770 | 7339 | 6366 | 5481 | 4844 | 4267 | 3906 | 3503 |
| 65–69 | 11 222 | 11 117 | 11 339 | 9835 | 9519 | 8540 | 7165 | 6565 | 5849 | 5498 |
| 70–74 | 14 196 | 14 495 | 13 833 | 13 909 | 12 360 | 12 402 | 10 896 | 9534 | 8868 | 8146 |
| 75–79 | 16 400 | 16 282 | 16 350 | 15 679 | 16 788 | 15 376 | 15 223 | 14 444 | 12 711 | 12 473 |
| 80–84 | 15 055 | 15 177 | 15 953 | 16 500 | 16 686 | 18 936 | 18 243 | 19 207 | 18 257 | 17 349 |
| 85–89 | 10 049 | 10 340 | 11 280 | 13 157 | 14 205 | 15 171 | 18 540 | 19 308 | 20 733 | 21 459 |
| 90–94 | 4166 | 4400 | 5427 | 6975 | 7965 | 9123 | 10 890 | 13 512 | 15 069 | 16 910 |
| 95–99 | 882 | 961 | 1393 | 2035 | 2516 | 3057 | 4144 | 4167 | 6208 | 6665 |
| 100–104 | 70 | 79 | 140 | 245 | 335 | 447 | 817 | 671 | 1073 | 1229 |
| 105–109 | 1 | 2 | 4 | 8 | 13 | 20 | 47 | 31 | 52 | 65 |
| 110–114 | 0 | 0 | 0 | 0 | 0 | 0 | 0 | 0 | 0 | 1 |
| 115–119 | 0 | 0 | 0 | 0 | 0 | 0 | 0 | 0 | 0 | 0 |

**Italy mortality 1974–2019, Male**

| Age interv. [y] | **1974** | **1979** | **1984** | **1989** | **1994** | **1999** | **2004** | **2009** | **2014** | **2019** |
|---|---|---|---|---|---|---|---|---|---|---|
| 0–4 | 2412 | 1742 | 1206 | 951 | 791 | 590 | 408 | 350 | 323 | 305 |
| 5–9 | 144 | 113 | 98 | 77 | 81 | 59 | 43 | 42 | 34 | 33 |
| 10–14 | 125 | 114 | 90 | 81 | 84 | 67 | 48 | 44 | 40 | 38 |
| 15–19 | 183 | 154 | 139 | 125 | 136 | 123 | 96 | 86 | 65 | 58 |
| 20–24 | 224 | 176 | 153 | 158 | 150 | 147 | 104 | 98 | 80 | 75 |
| 25–29 | 251 | 215 | 179 | 182 | 220 | 161 | 122 | 106 | 97 | 92 |
| 30–34 | 369 | 284 | 258 | 242 | 277 | 226 | 160 | 144 | 133 | 128 |
| 35–39 | 499 | 462 | 380 | 379 | 361 | 316 | 256 | 222 | 224 | 208 |
| 40–44 | 798 | 684 | 662 | 570 | 530 | 488 | 402 | 376 | 352 | 349 |
| 45–49 | 1271 | 1131 | 1011 | 981 | 829 | 795 | 668 | 620 | 581 | 550 |
| 50–54 | 1865 | 1778 | 1576 | 1429 | 1316 | 1178 | 1065 | 1018 | 906 | 828 |
| 55–59 | 3033 | 2583 | 2437 | 2172 | 1928 | 1783 | 1600 | 1513 | 1395 | 1322 |
| 60–64 | 4331 | 4345 | 3770 | 3408 | 3085 | 2697 | 2453 | 2327 | 2160 | 2063 |
| 65–69 | 6860 | 6461 | 6351 | 5375 | 4913 | 4439 | 3745 | 3580 | 3299 | 3124 |
| 70–74 | 10 825 | 10 375 | 9546 | 9249 | 7885 | 7280 | 6167 | 5722 | 5270 | 5000 |
| 75–79 | 16 790 | 15 818 | 14 897 | 13 451 | 13 604 | 11 735 | 10 410 | 9881 | 8690 | 8573 |
| 80–84 | 20 142 | 19 922 | 20 321 | 19 358 | 18 257 | 19 401 | 16 351 | 16 591 | 15 252 | 14 569 |
| 85–89 | 17 798 | 18 907 | 20 624 | 20 657 | 21 659 | 21 525 | 23 489 | 22 727 | 22 398 | 22 507 |
| 90–94 | 9557 | 11 251 | 12 428 | 14 844 | 16 353 | 17 830 | 19 820 | 22 284 | 22 334 | 23 485 |
| 95–99 | 2332 | 3168 | 3522 | 5492 | 6470 | 7696 | 9924 | 9947 | 13 000 | 13 014 |
| 100–104 | 187 | 310 | 345 | 785 | 1021 | 1380 | 2480 | 2182 | 3150 | 3404 |
| 105–109 | 4 | 8 | 9 | 33 | 49 | 79 | 187 | 139 | 215 | 271 |
| 110–114 | 0 | 0 | 0 | 0 | 1 | 1 | 3 | 2 | 3 | 4 |
| 115–119 | 0 | 0 | 0 | 0 | 0 | 0 | 0 | 0 | 0 | 0 |

**Italy mortality 1974–2019, Female**